%% file: main.tex
\documentclass[]{externalized}

\input{common}

\title{Externalization in LLM Agents: A Unified Review of \\ Memory, Skills, Protocols and Harness Engineering}

\author[1]{Chenyu Zhou}
\author[1,*]{Huacan Chai}
\author[1,*]{Wenteng Chen}
\author[2,3,*]{Zihan Guo}
\author[1,*]{Rong Shan}
\author[1,*]{Yuanyi Song}
\author[1,*]{Tianyi Xu}
\author[1,*]{Yingxuan Yang}
\author[1,*]{Aofan Yu}
\author[1,*]{Weiming Zhang}
\author[1,*]{Congming Zheng}
\author[1,*]{Jiachen Zhu}
\author[4]{Zeyu Zheng}
\author[1]{Zhuosheng Zhang}
\author[5]{Xingyu Lou}
\author[5]{Changwang Zhang}
\author[5]{Zhihui Fu}
\author[5,\dag]{Jun Wang}
\author[1,\dag]{Weiwen Liu}
\author[1,\dag]{Jianghao Lin}
\author[1,3,\dag]{Weinan Zhang}

\affiliation[1]{Shanghai Jiao Tong University}
\affiliation[2]{Sun Yat-Sen University}
\affiliation[3]{Shanghai Innovation Institute}
\affiliation[4]{Carnegie Mellon University}
\affiliation[5]{OPPO}
\contribution[*]{Equal contribution}
\contribution[\dag]{Corresponding authors}
\correspondence{\email{\{wwliu, linjianghao, wnzhang\}@sjtu.edu.cn}, \email{junwang.lu@gmail.com}}

\abstract{
\input{text/00_abstract}
}

\date{\today}

\begin{document}
\maketitle
\newpage
\tableofcontents
\newpage

\input{text/01_introduction}
\input{text/02_background}
\input{text/03_memory}
\input{text/04_skills}

\input{text/05_protocols}
\input{text/06_harness}
\input{text/07_cross_cutting}
\input{text/08_future_discussion}
\input{text/09_conclusion}

\newpage

\bibliographystyle{abbrvnat}
\bibliography{cite}

\end{document}

%% file: common.tex
\usepackage[T1]{fontenc}
\usepackage[utf8]{inputenc}
\usepackage{microtype}
\usepackage{inconsolata}
\usepackage{graphicx}
\usepackage{hyperref}
\usepackage{url}
\usepackage{booktabs}
\usepackage{amsfonts}
\usepackage{amsmath}
\usepackage{amssymb}
\usepackage{amsthm}
\usepackage{mathtools}
\usepackage{enumitem}
\usepackage{multirow}
\usepackage{subcaption}
\usepackage{xcolor}
\usepackage{listings}
\usepackage{natbib}
\usepackage{multicol}
\usepackage{pifont}
\usepackage{xspace}
\usepackage{wrapfig}
\AtBeginDocument{%
  \providecommand\BibTeX{{%
    \normalfont B\kern-0.5em{\scshape i\kern-0.25em b}\kern-0.8em\TeX}}}

\makeatletter
\DeclareRobustCommand\onedot{\futurelet\@let@token\@onedot}
\def\@onedot{\ifx\@let@token.\else.\null\fi}

\makeatother

\definecolor{codegreen}{rgb}{0,0.6,0}
\definecolor{codegray}{rgb}{0.5,0.5,0.5}
\definecolor{codepurple}{rgb}{0.58,0,0.82}
\definecolor{backcolour}{rgb}{0.95,0.95,0.92}

\lstdefinestyle{mystyle}{
  backgroundcolor=\color{backcolour},
  commentstyle=\color{codegreen},
  keywordstyle=\color{magenta},
  numberstyle=\tiny\color{codegray},
  stringstyle=\color{codepurple},
  basicstyle=\ttfamily\footnotesize,
  breakatwhitespace=false,
  breaklines=true,
  captionpos=b,
  keepspaces=true,
  numbers=left,
  numbersep=5pt,
  showspaces=false,
  showstringspaces=false,
  showtabs=false,
  tabsize=2
}

%% file: text/00_abstract.tex
Large language model (LLM) agents are increasingly built less by changing model weights than by reorganizing the runtime around them. Capabilities that earlier systems expected the model to recover internally are now externalized into memory stores, reusable skills, interaction protocols, and the surrounding harness that makes these modules reliable in practice. This paper reviews that shift through the lens of externalization. Drawing on the idea of cognitive artifacts, we argue that agent infrastructure matters not merely because it adds auxiliary components, but because it transforms hard cognitive burdens into forms that the model can solve more reliably. Under this view, memory externalizes state across time, skills externalize procedural expertise, protocols externalize interaction structure, and harness engineering serves as the unification layer that coordinates them into governed execution. We trace a historical progression from weights to context to harness, analyze memory, skills, and protocols as three distinct but coupled forms of externalization, and examine how they interact inside a larger agent system. We further discuss the trade-off between parametric and externalized capability, identify emerging directions such as self-evolving harnesses and shared agent infrastructure, and discuss open challenges in evaluation, governance, and the long-term co-evolution of models and external infrastructure. The result is a systems-level framework for explaining why practical agent progress increasingly depends not only on stronger models, but on better external cognitive infrastructure.

%% file: text/01_introduction.tex
\section{Introduction}
\label{sec:introduction}

\begin{figure}[!htbp]
\centering
\includegraphics[width=\linewidth]{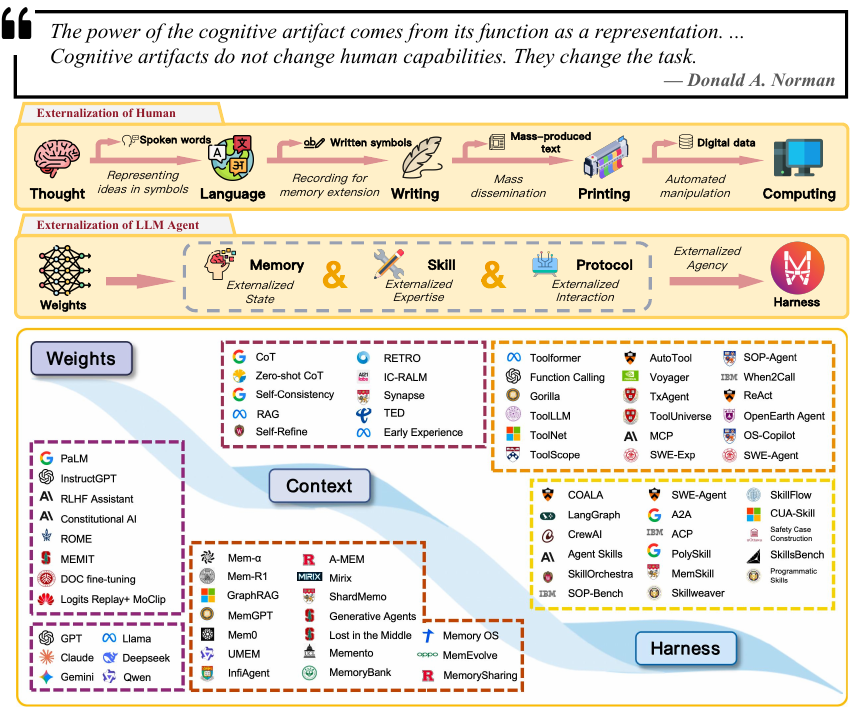}
\caption{\textbf{Externalization as the organizing principle of LLM agent design.} \emph{Upper panel:} The arc of human cognitive externalization from thought through language, writing, printing, to digital computation. \emph{Middle panel:} The corresponding externalization arc for LLM agents, from weights through three externalization dimensions---Memory (externalized state), Skills (externalized expertise), and Protocols (externalized interaction)---to the Harness that unifies them. \emph{Lower panel:} A literature landscape mapping representative works onto three capability layers---Weights, Context, and Harness---illustrating how research threads have progressively migrated outward. The parallel between the two arcs encodes a recursive claim: LLM agents achieve reliable agency by externalizing cognitive burdens along the same representational dimensions that have driven human cognitive history.}
\label{fig:opening}
\end{figure}

The history of human civilization can also be read as a history of cognitive externalization. Spoken language transformed private thought into shareable symbolic form.
Writing moved knowledge from fragile biological memory into persistent material records.
Printing mechanized the reproduction of knowledge at social scale.
Digital computation relocated arithmetic and symbolic manipulation from neural labor to programmable machines.
Across these transitions, the critical change was not that humans became less capable without the artifact. Rather, the artifact reorganized the cognitive system by shifting selected burdens outward and freeing limited internal resources for planning, abstraction, and creativity \citep{norman1993things}. The same pattern of outward delegation now recurs at the frontier of machine intelligence, in the design of large language model agents.

This perspective has a natural theoretical anchor in the idea of cognitive artifacts \citep{norman1991cognitive,norman1993things}. The central insight is that external aids do not merely amplify an unchanged internal ability; they often transform the task itself. A shopping list does not expand biological memory capacity. It changes a difficult recall problem into a recognition problem. A map does not simply make navigation ``stronger.'' It converts hidden spatial relations into visible structure. The power of an artifact therefore lies in representational transformation: it restructures the problem so that the agent can solve it more reliably with the competencies it already has \citep{norman1991cognitive}.

We argue that the same logic now governs the most consequential design choices in LLM-based agents. Our central thesis is that externalization---the progressive relocation of cognitive burdens from the model's internal computation into persistent, inspectable, and reusable external structures---is the transition logic---the mechanism that explains why each architectural shift has occurred and what forms of reliability it sought to preserve---that unifies recent advances in memory, skills, protocols, and harness engineering for language agents. This is not merely a claim about engineering convenience. It is a claim about where reliable agency comes from: not from ever-larger models alone, but from the systematic restructuring of task demands so that internal capabilities and external infrastructure jointly cover the full range of competencies required \citep{norman1991cognitive,sumers2024coala}.

Figure~\ref{fig:opening} summarizes the argument. The upper panel traces the familiar arc of human cognitive externalization; the middle panel presents the corresponding arc for LLM agents, from weights through three externalization dimensions---memory, skills, and protocols---to the harness that unifies them; the lower panel maps the resulting literature landscape onto three capability layers---Weights, Context, and Harness. Figure~\ref{fig:harness_architecture} complements this view with an architectural overview of the externalized agent, showing the harness at the center with the three externalization dimensions and their operational elements orbiting it. Memory externalizes state across time, skills externalize procedural expertise, and protocols externalize interaction structure. The parallel between the two arcs encodes a recursive claim: LLM agents are themselves artifacts operating inside the latest major human externalization, digital computation. The common mechanism is representational transformation in Norman's sense \citep{norman1991cognitive}: recall becomes recognition, improvised generation becomes composition, and ad hoc coordination becomes structured contract.


This lens is especially clarifying for understanding current practice. Contemporary progress is often narrated as a race for larger models, better training procedures, or more sophisticated reasoning traces. Those factors matter, but they do not fully explain the pattern observed in practical systems. Many of the largest gains in reliability do not come from changing the base model at all. They come from changing the environment around the model: adding persistent memory, organizing reusable skills, standardizing tool interfaces, constraining execution, instrumenting behavior, and routing work through explicit control logic \citep{sumers2024coala,wang2024survey,li2025review,luo2025agent}. In practice, the question is increasingly not only ``how capable is the model?'' but also ``what burdens have been externalized so the model no longer has to solve them internally every time?''

An unaided LLM still faces three recurrent mismatches that map directly onto the three harness dimensions. Its context window is finite and session memory is weak or absent, creating a continuity problem that memory externalization addresses. Long multi-step procedures are often rederived rather than executed consistently, creating a variance problem that skill externalization addresses. Interactions with external tools, services, and collaborators remain brittle when left to free-form prompting alone, creating a coordination problem that protocol externalization addresses \citep{sumers2024coala,packer2023memgpt}. Externalization matters because it turns each of these burdens into a form the model can handle more reliably.

A concrete example helps fix the intuition. Consider a software engineering agent asked to implement a feature in a large repository, run tests, and open a pull request. Without externalization, the model must keep repository structure, project conventions, workflow state, and tool interactions active inside a fragile prompt. With externalization, persistent project memory supplies context, reusable skill documents encode conventions and workflow, protocolized tool interfaces enforce correct schemas, and the harness sequences steps, validates outputs, and manages failures. The base model may remain unchanged; what changes is the representation of the task it is asked to solve.


This broader perspective also aligns with the intuition behind distributed and extended cognition: once crucial parts of remembering, guiding action, and coordinating interaction are delegated to external structures, intelligence is no longer localized in the model alone \citep{clark1998extended}. We draw on this tradition for its core engineering insight---that the boundary between ``agent'' and ``environment'' is a design choice with real performance consequences---rather than committing to its stronger ontological claims. Our focus is pragmatic: we treat externalization as a design principle whose value is measured by the reliability, composability, and governability of the resulting system.


We now turn to the three dimensions of externalization that constitute the harness, each corresponding to one of the representational transformations highlighted in Figure~\ref{fig:opening} (middle panel):

\paragraph{Memory systems externalize state across time.} Rather than relying on the context window as the sole carrier of history, memory systems allow accumulated knowledge---user preferences, prior trajectories, resolved ambiguities, domain facts---to persist beyond any single session and be selectively retrieved when relevant. The core transformation is from recall to recognition: the agent no longer needs to regenerate past knowledge from latent weights; it retrieves it from a persistent, searchable store \citep{lewis2020retrieval,park2023generative,packer2023memgpt,chhikara2025mem0,xu2025amem}.

\paragraph{Skill systems externalize procedural expertise.} Rather than relying on the model's weights to regenerate task-specific know-how on every invocation, skill systems package procedures, best practices, and operating guidance into reusable artifacts. The core transformation is from generation to composition: the agent assembles behavior from pre-validated components rather than improvising each step de novo \citep{openai2023function,schick2023toolformer,wang2023voyager,anthropic2025skills,anthropic2026skillsdocs,jiang2026agenticskills}.

\paragraph{Protocols externalize interaction structure.} Rather than relying on ad hoc prompt-level coordination with tools, services, and other agents, protocols define explicit machine-readable contracts for discovery, invocation, delegation, and permission management. The core transformation is from ad-hoc to structured: ambiguous, fragile communication becomes interoperable, governable exchange \citep{anthropic2024mcp,google2025a2a,ehtesham2025protocols}.

The harness is the engineering layer that hosts all three dimensions and provides the orchestration logic, constraints, observability, and feedback loops that make externalized cognition cohere in practice. It is not a fourth kind of externalization alongside memory, skills, and protocols. It is the runtime environment within which these forms of externalization operate and interact.


These dimensions do not evolve in isolation. Memory expansion can compete with skill loading for scarce context budget. Protocol standardization can improve interoperability while constraining how capabilities are packaged and invoked. Skill execution generates traces that later become memory, and memory retrieval can influence which skills and protocol paths are chosen next. The harness must mediate all of these interactions. We preview these system-level couplings here and analyze them in detail in Section~\ref{sec:cross_cutting}.


These directions have each developed substantial technical ecosystems. Memory research has progressed from simple retrieval augmentation to more selective and tiered memory architectures \citep{lewis2020retrieval,packer2023memgpt,chhikara2025mem0,xu2025amem}. Skill-related work has expanded from narrow function calling and tool learning toward reusable capability packages, registries, and progressive disclosure mechanisms \citep{openai2023function,schick2023toolformer,wang2023voyager,anthropic2025skills,anthropic2026skillsdocs,jiang2026agenticskills}. Protocol work has moved from custom tool schemas and framework-specific glue code toward more standardized interface layers for agent-tool and agent-agent interaction \citep{anthropic2024mcp,google2025a2a,ehtesham2025protocols}. Existing surveys illuminate important slices of this landscape, including retrieval-augmented generation \citep{gao2024rag}, deep search~\cite{xi2025survey}, tool learning and use \citep{qu2024tool}, broad agent architectures \citep{wang2024survey,li2025review,luo2025agent}, and protocol interoperability \citep{ehtesham2025protocols}. The closest conceptual bridge is CoALA \citep{sumers2024coala}. What remains underdeveloped is a common account of why these developments are converging as forms of externalization and how that convergence reshapes the definition of an agent.


Our goal is therefore not to provide another component-level survey in isolation, nor to reduce agent progress to one specific framework. Instead, we offer a systems-level review organized around four claims:

\begin{itemize}[leftmargin=1.5em]
\item Memory systems externalize an agent's state across time and convert long-horizon continuity into selective retrieval.
\item Skill systems externalize procedural expertise and convert implicit know-how into explicit reusable operating guidance.
\item Protocols externalize interaction structure and convert ambiguous communication into interoperable, machine-readable contracts.
\item Harness engineering unifies these externalized modules into a coherent runtime environment with constraints, observability, feedback loops, and control points.
\end{itemize}

The remainder of the paper proceeds as follows. Section~\ref{sec:background} traces the historical path from weights to context to harness. Sections~\ref{sec:memory} through~\ref{sec:protocols} analyze memory, skills, and protocols as three distinct but complementary forms of externalization. Section~\ref{sec:harness} presents harness engineering as the integrative discipline of externalized agent design, and Section~\ref{sec:cross_cutting} examines the main cross-cutting interactions among the modules. Section~\ref{sec:future_discussion} discusses future directions toward more adaptive and self-evolving forms of externalization, and Section~\ref{sec:conclusion} concludes with broader implications for agent research.

%% file: text/02_background.tex
\section{Background: From Weights to Context to Harness}
\label{sec:background}

\begin{figure}[!htbp]
\centering
\includegraphics[width=0.95\linewidth]{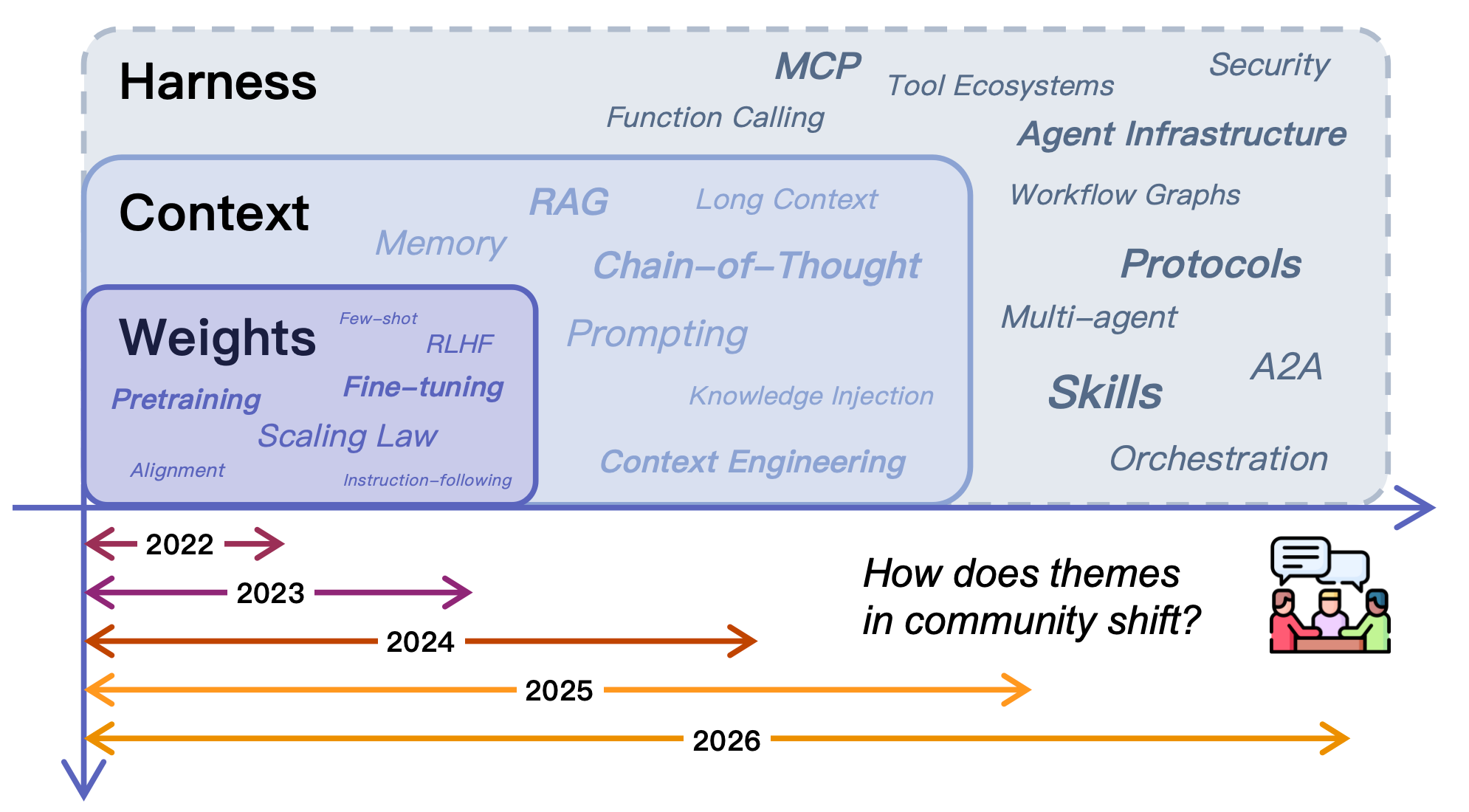}
\caption{\textbf{Community theme evolution across three capability layers.} The stacked layers---Weights, Context, and Harness---show how the center of gravity in the LLM agent community has shifted outward over time, from parametric knowledge and prompting toward harness-level infrastructure such as tool ecosystems, protocols, skills, and multi-agent orchestration.}
\label{fig:evolve}
\end{figure}

The recent history of LLM agents can be understood as a progressive movement outward from the model itself. Capabilities were first treated as properties of weights, then as properties of prompts and context windows, and are now increasingly treated as properties of the broader infrastructure in which the model operates. Figure~\ref{fig:evolve} visualizes this trajectory as three stacked layers---Weights, Context, and Harness---unfolding across a timeline from 2022 to 2026, illustrating how research themes in the community have shifted over time. The lower panel of Figure~\ref{fig:opening} complements that view with a literature landscape, mapping representative works to the three layers. The stages are layered rather than mutually exclusive---weights remain important even in the most infrastructure-heavy systems---but each stage changes where developers place the system's mutable intelligence and, consequently, where they invest most of their engineering effort.

\subsection{Capability in Weights}
\label{sec:bg_weights}

The \textbf{Weights} layer in Figures~\ref{fig:evolve} and~\ref{fig:opening} corresponds to the earliest wave of modern LLM deployment, in which capability was identified almost entirely with model parameters. Pretraining on large corpora compressed broad statistical regularities, world knowledge, and latent reasoning habits into the weights \citep{brown2020language,chowdhery2023palm,touvron2023llama}. Scaling laws revealed predictable relationships between parameter count, data volume, and loss, reinforcing the intuition that progress tracked directly with model size \citep{kaplan2020scaling,hoffmann2022training}. By the time systems such as GPT-4 \citep{openai2023gpt4}, Gemini \citep{team2023gemini}, DeepSeek-V3 \citep{deepseek2025v3}, and Qwen2.5 \citep{qwen2025qwen25} demonstrated broad multi-task competence, the dominant narrative in much of the field equated better agents with bigger, better-trained models. Supervised fine-tuning and preference optimization then shaped these models into more useful assistants by teaching instruction following, conversational style, refusal behavior, and domain-specific conventions \citep{ouyang2022training,bai2022training}; direct preference optimization further simplified this alignment stage by eliminating the need for a separate reward model \citep{rafailov2023direct}. From this viewpoint, improvement largely meant modifying or replacing the model itself.

This paradigm remains foundational, and weight-space capability offers several advantages: fast inference without external lookups, compact deployment, and strong generalization across many tasks without task-specific plumbing. The same model that answers a medical question can write a poem, debug a program, or summarize a legal contract, all without any change to the surrounding system. For one-shot, context-contained tasks, the weight-centric view is often sufficient.

However, weight-space encoding also couples knowledge, procedure, and policy too tightly to a static artifact. Updating a single fact---say, the current head of state of a country---requires retraining, knowledge editing \citep{meng2022locating,mitchell2022fast,yao2023editing}, or patching through additional alignment layers, all of which risk unintended side effects on other capabilities. Auditing why a model behaved a certain way is difficult because relevant knowledge is distributed across billions of parameters rather than encoded as inspectable modules \citep{zhao2024explainability}. Personalization is also awkward: a single set of weights is asked to serve millions of users with different histories, preferences, and constraints, yet it has no mechanism to differentiate among them at the parameter level.

A central limitation of parametric knowledge is that it is difficult to selectively update, compose, and govern. As long as agents were confined to single-turn question answering, these weaknesses were often manageable. As systems moved into long-horizon task execution---where state accumulates, procedures must be followed reliably, and multiple tools must be coordinated---the difficulty of modularly managing knowledge, skills, and interaction rules inside the weights became more operationally salient. This shift encouraged developers to relocate some of these burdens into the next layer rather than relying on the model parameters alone.

\begin{figure*}[!t]
\centering
\includegraphics[width=0.9\textwidth]{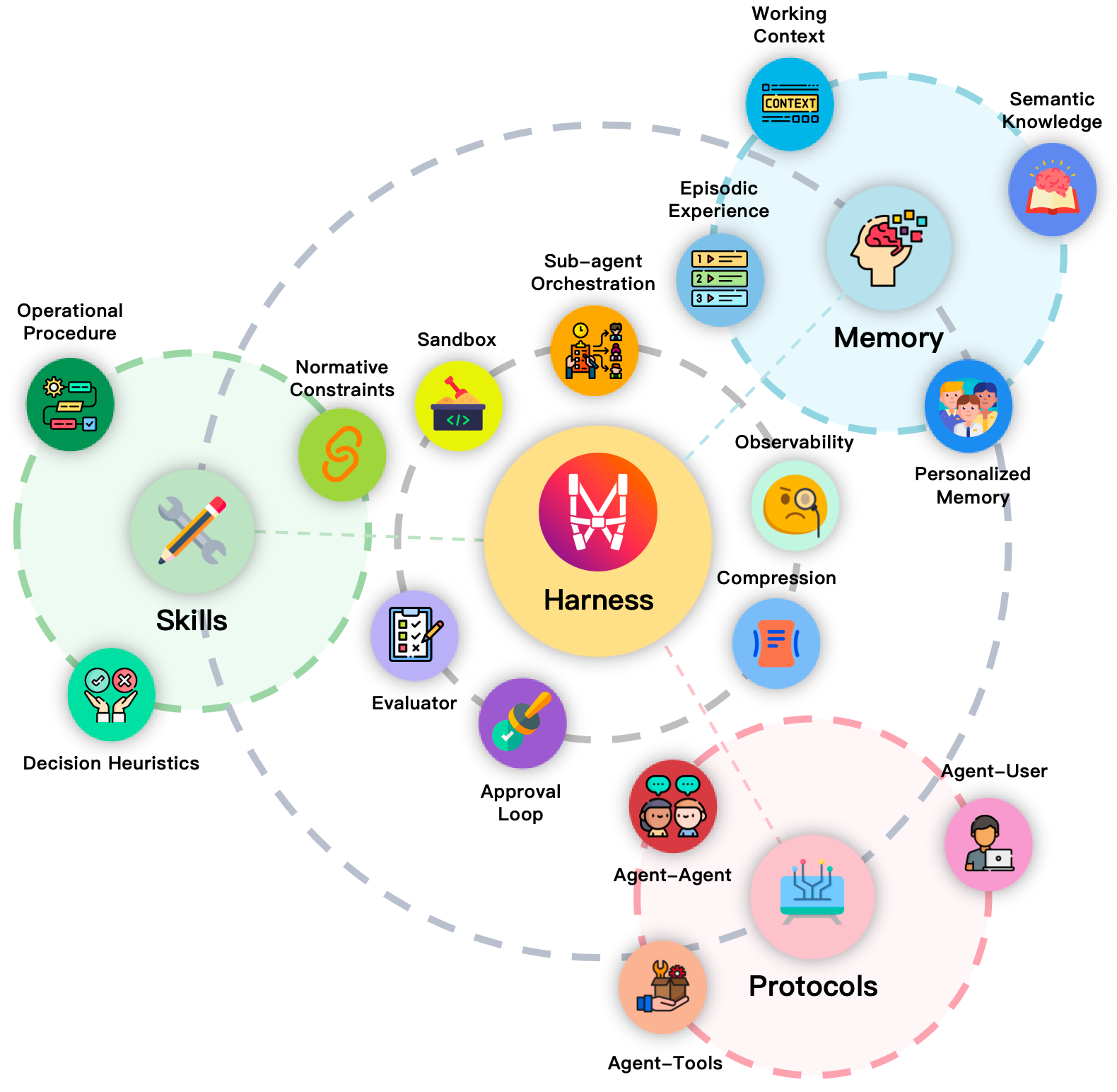}
\caption{\textbf{Externalization architecture of a harnessed LLM agent.} The Harness sits at the center; three externalization dimensions---\textbf{Memory} (working context, semantic knowledge, episodic experience, personalized memory), \textbf{Skills} (operational procedures, decision heuristics, normative constraints), and \textbf{Protocols} (agent--user, agent--agent, agent--tools)---orbit around it. Operational elements such as sandboxing, observability, compression, evaluation, approval loops, and sub-agent orchestration mediate the interaction between the harness core and the externalized modules.}
\label{fig:harness_architecture}
\end{figure*}

\subsection{Capability in Context}
\label{sec:bg_context}

The \textbf{Context} layer represents the stage at which attention shifted from model modification to input design. Prompt engineering demonstrated that model behavior could be substantially altered without touching the weights: few-shot examples, role descriptions, chain-of-thought decomposition, and self-consistency traces all changed how the same model performed on the same underlying task \citep{wei2022chain,wang2023selfconsistency,kojima2022large}. Techniques for more structured reasoning soon followed. ReAct interleaved reasoning traces with tool actions in a single generation loop, showing that prompting alone could produce agent-like behavior without any architectural change \citep{yao2023react}. Tree of Thoughts generalized chain-of-thought into deliberate search over intermediate reasoning states \citep{yao2024tree}. Self-Refine introduced iterative self-critique, demonstrating that models could improve their own outputs through multi-turn prompting loops \citep{madaan2023selfrefineiterativerefinementselffeedback}. Automatic prompt optimization further reduced the manual burden by using the model itself to search over the prompt space \citep{zhou2023large,pryzant2023automatic}. Retrieval-augmented generation (RAG) introduced a more systematic form of externalization by dynamically injecting external documents into the context at query time \citep{lewis2020retrieval,borgeaud2022improving,ram2023incontext,gao2024rag}. Attention thus shifted from what the model had internalized to the information pipeline surrounding each invocation.

This stage made agent design substantially more flexible. Developers could attach local instructions, domain knowledge, output schemas, and retrieved evidence at runtime without any gradient update. Context became the medium through which developers staged cognition for the model---a working surface on which the right information could be assembled just before the model needed it. In many practical systems, iterating on prompts and retrieval pipelines proved substantially cheaper and faster than fine-tuning. The model could remain frozen while the surrounding prompt template, retrieval logic, and tool specification evolved rapidly.

The context-centric stage can also be interpreted through Norman's notion of representational transformation. A difficult recall problem---``does the model know fact $X$?''---was converted into a recognition problem: ``given that fact $X$ has been placed in context, can the model use it?'' This resembles the recall-to-recognition shift associated with writing in the human externalization arc (Figure~\ref{fig:opening}, upper panel). The model did not need to have memorized the answer; it needed only to recognize and apply the relevant passage once it was provided. In Figure~\ref{fig:evolve}, this transition corresponds to the emergence of prompting, RAG, chain-of-thought, and related techniques in the Context layer~\cite{zhang2024agentic,ram2023incontext}.

Context-centric design also has important constraints. Context windows are finite, costly at scale, and often noisy when overloaded with marginally relevant material. Long prompts can degrade performance rather than improve it: the ``lost in the middle'' phenomenon shows that models attend unevenly across long inputs, with retrieval accuracy dropping sharply for information placed in the center of the context \citep{liu2024lost}. Even as context lengths have expanded dramatically---from 2K tokens to over 100K and beyond \citep{chen2023extending,peng2024yarn}---the fundamental tension persists: more capacity does not eliminate the need for selective curation. Context is also ephemeral: unless state is explicitly externalized elsewhere, every new session begins with partial amnesia. As systems become more complex, prompt assembly alone can become a brittle and ad hoc control mechanism. A model can be given more instructions, but that does not mean the system knows how to persist state across sessions, schedule multi-step workflows, coordinate among sub-agents, recover from partial failures, or enforce behavioral constraints over time. These limitations help explain the next outward step.

\subsection{Capability through Infrastructure}
\label{sec:bg_infra}

The \textbf{Harness} layer---the topmost band in Figure~\ref{fig:evolve} and the rightmost region in Figure~\ref{fig:opening} (lower panel)---represents the current stage, in which capability extends beyond prompt management into persistent infrastructure. As context windows became saturated and prompt templates more unwieldy, engineering attention increasingly shifted from ``what should we tell the model?'' to ``what environment should the model operate in?'' In mature agent systems, reliability increasingly depends on external memory stores, tool registries, protocol definitions, sandboxes, sub-agent orchestration, compression pipelines, evaluators, test harnesses, and approval loops \citep{wang2024survey,li2025review,luo2025agent,xi2023rise}.

The earliest manifestations of this shift were simple but revealing. Projects such as Auto-GPT \citep{richards2023autogpt} and BabyAGI \citep{nakajima2023babyagi} wrapped an LLM in a loop with a task queue, persistent memory, and web access, showing that even a minimal harness could sustain behavior that no single prompt could. More principled frameworks quickly followed: AutoGen formalized multi-agent message exchange \citep{wu2023autogen}, MetaGPT added role-based collaboration and explicit procedures \citep{hong2023metagpt}, CAMEL explored structured dialogue for task decomposition \citep{li2023camel}, and Reflexion persisted feedback across episodes \citep{shinn2023reflexion}. Across these systems, the common move was to shift burden out of the model and into surrounding structure.

The same move is now visible across deployment domains. Coding agents embed the model in development harnesses with files, shells, version control, tests, and reusable skill artifacts; SWE-agent and OpenHands are representative examples \citep{yang2024sweagent,wang2024opendevin}. Research and enterprise agents add retrieval, approvals, browsing, and long-horizon orchestration pipelines, as in Deep Research-style systems \citep{openai2025deepresearch,google2025geminiresearch}. Embodied and workflow systems such as Voyager, LangGraph, CrewAI, and OS-Copilot likewise make control flow, environment access, and reuse explicit \citep{wang2023voyager,langchain2024langgraph,crewai2024,wu2024oscopilot}. The recurring pattern is that reliability problems are increasingly solved by changing the environment rather than by prompting alone.

As shown in Figure~\ref{fig:harness_architecture}, the harness encompasses three major classes of externalization---memory, skills, and protocols---which correspond to the three major classes of burden that the harness absorbs. Memory systems externalize state across time, so that continuity no longer depends on ephemeral context. Skill systems externalize procedural expertise, so that complex workflows are loaded rather than reinvented. Protocols externalize interaction structure, so that tool and agent coordination follows governed contracts rather than ad hoc prompting. Together, these elements make up the harness: the persistent infrastructure that envelops the model and transforms the tasks it faces into forms that its internal competencies can handle more reliably.

Under this framing, ``agent engineering'' increasingly takes the form of ``harness engineering.'' The model remains the core reasoning engine, but it is no longer the sole location of intelligence. Capability is distributed across the structures that shape what the model sees, remembers, calls, and is allowed to do.

\subsection{Externalization as the Transition Logic}
\label{sec:bg_transition}

Taken together, the path from weights to context to harness is a story of externalization in Norman's sense \citep{norman1991cognitive}: burdens that are hard to manage inside the model are progressively moved into explicit artifacts outside it, and the task seen by the model is correspondingly transformed. Mutable knowledge moves from weights into retrieval systems and runtime context, converting recall into recognition. Reusable procedures move from implicit habits into explicit skills, converting improvised generation into structured composition. Interaction rules move from ad hoc prompting into protocols, converting ambiguous coordination into governed contracts. Runtime reliability, in turn, moves into harness logic, where constraints, observability, and feedback loops can be made explicit.

This redistribution is best understood as a response to mismatch. LLMs are strong at flexible synthesis and reasoning over provided information; they are less reliable at stable long-term memory, procedural repeatability, and governed interaction with external systems. Externalization therefore constructs a larger cognitive system around the model rather than replacing it, a view consistent with cognitive-architecture accounts such as CoALA \citep{sumers2024coala}. The three harness dimensions follow directly from this framing: memory addresses continuity over time, skills address consistency of procedure, and protocols address structure of interaction. The following sections examine each in detail.

%% file: text/03_memory.tex
\section{Externalized State: Memory}
\label{sec:memory}


Memory externalization addresses the temporal burden of agency. A bare language model must carry continuity, prior experience, user-specific facts, and partially completed work inside an ephemeral prompt. Once tasks extend across sessions, branches, or interruptions, that burden becomes both unstable and expensive. Memory externalizes it into persistent state that can be written, updated, and retrieved outside the model.

In harnessed agents, memory is more than an archive. It supplies checkpoints for resumable execution, traces from which skills can be distilled, statistics that influence protocol routing, and persistent state that governance mechanisms can inspect and constrain. To make that role precise, this section asks three linked questions: what burden memory externalizes, how the design space has evolved, and how memory couples to the broader harness. Section~\ref{memory: what} clarifies which kinds of state are externalized; Section~\ref{memory: how} surveys the main architectural choices; Section~\ref{memory: harness} turns to the demands imposed by harnessed agent systems; and Section~\ref{memory: cognitive artifact} closes the chapter by interpreting memory through the lens of cognitive artifacts.

\begin{figure}[!htbp]
  \centering
  \includegraphics[width=\linewidth]{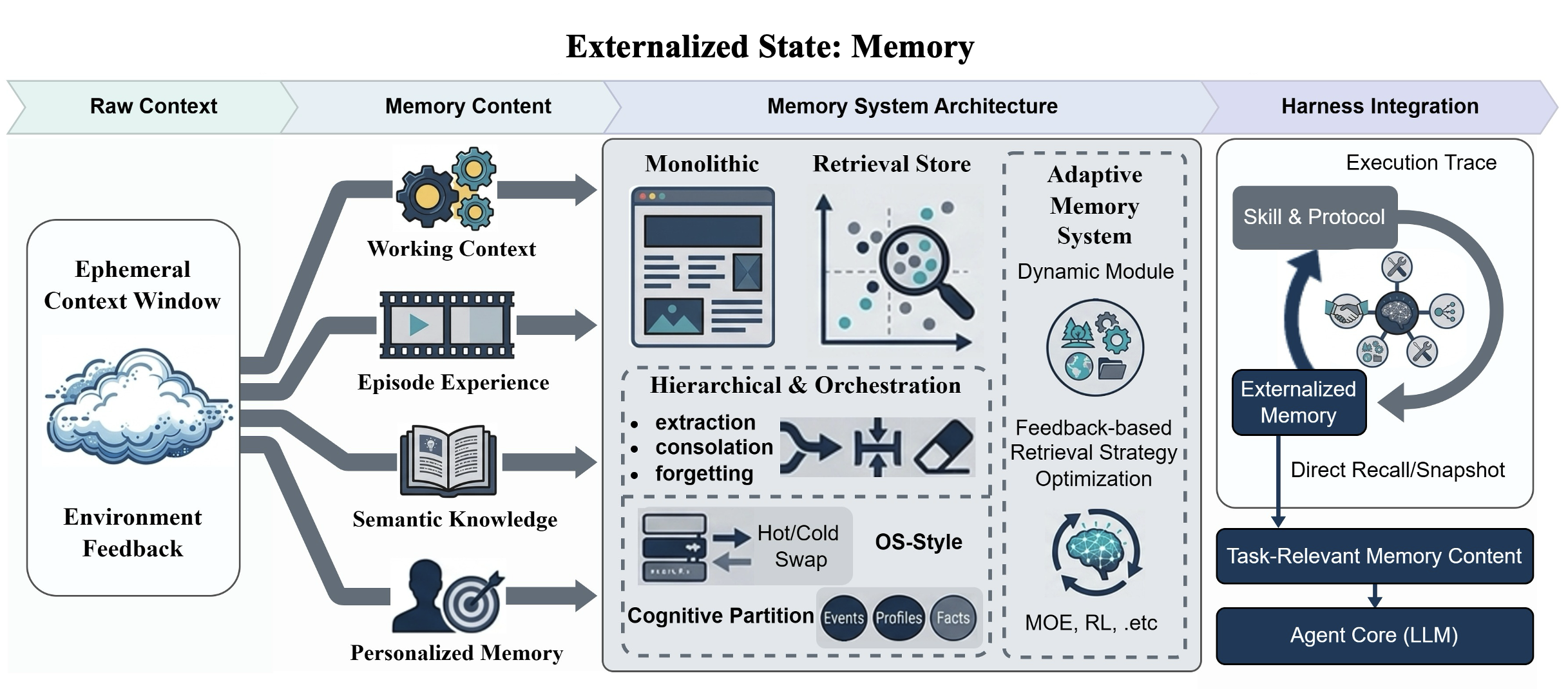}
  \caption{\textbf{Memory as externalized state.} Raw context from the ephemeral context window and environment feedback is converted into four persistent memory dimensions---working context, episodic experience, semantic knowledge, and personalized memory. These dimensions are organized through progressively more managed architectures: monolithic context, retrieval stores, hierarchical orchestration (with extraction, consolidation, forgetting, and OS-style hot/cold swapping), and adaptive memory systems (with dynamic modules and feedback-based strategy optimization via MOE, RL, etc.). On the harness side, execution traces from skills and protocols flow into externalized memory, which in turn supplies task-relevant content back to the agent core through direct recall and curated snapshots.}
  \label{fig:memory_externalization}
\end{figure}

\subsection{What Is Externalized: The Content of State}
\label{memory: what}

The essence of memory lies in decoupling the agent's state across time from its transient context. The relevant contents are not every external artifact in the harness, but the records that preserve continuity: current task state, past execution experiences, abstracted knowledge, and persistent user or environment context. To maintain coherent behavior across long-horizon interactions, the memory system must categorize and manage these records according to their temporal properties and retrieval needs. Drawing inspiration from classical taxonomies of human memory and adapting them to LLM agents, we distinguish the following four dimensions of externalized state:

\paragraph{Working context.} Working context is the live intermediate state of the current task: open files, temporary variables, active hypotheses, partial plans, and execution checkpoints. It changes quickly and loses value if it is stale, but without externalization it disappears as soon as the context window resets or a process is interrupted. Coding agents illustrate the point well. By materializing drafts, terminal state, and workspace artifacts outside the prompt, systems such as OpenHands and SWE-style agents can resume from the current operating state rather than reconstructing it from scratch~\citep{wang2025openhands, yang2024swe}.

\paragraph{Episodic experience.} Episodic experience records what happened in prior runs: decision points, tool calls, failures, outcomes, and reflections. Its value is not merely archival. Retrieved episodes can serve as concrete precedents, help the agent avoid repeating known mistakes, and supply raw material for later abstraction. Reflexion made this pattern explicit by storing reflective summaries from failed attempts as reusable experience~\citep{shinn2023reflexion}. AriGraph extends the idea further by treating local interaction trajectories in unfamiliar environments as episodic memory from which a broader world model can be built~\citep{anokhin2024arigraph}.

\paragraph{Semantic knowledge.} Semantic knowledge stores abstractions that outlive any single episode: domain facts, general heuristics, project conventions, and stable world knowledge. Unlike episodic memory, it is not organized around a specific time and place~\citep{li2024memory, de2022rethinking}. The difference is not only granularity but function. Episodic memory says what happened in a case; semantic memory says what tends to hold across cases. In current systems, knowledge bases and Retrieval-Augmented Generation (RAG) corpora are the most common form of externalized semantic memory. The longer-term trend is more ambitious: agents increasingly try to distill semantic guidance from accumulated trajectories rather than relying only on static human-authored documents.

\paragraph{Personalized memory.} Personalized memory tracks stable information about particular users, teams, or environments: preferences, habits, recurring constraints, and prior interactions. This state should not be collapsed into the agent's general self-improvement store, because user-specific traces obey different retention, retrieval, and privacy rules~\cite{xi2024memocrs,lin2025can}. Recent systems make this separation explicit. IFRAgent builds a repository of user habits from demonstrations in mobile environments~\citep{wu2025quick}; web agents use externalized profiles to infer implicit preferences~\citep{cai2025large}; and conversational systems such as VARS store cross-session preference cards in isolated user memory spaces~\citep{hao2026user}. Personalized memory is therefore the layer that lets an agent adapt over time without confusing long-term user modeling with general task knowledge.

These four layers do not exhaust everything that may later become useful to the agent. Repeated procedural regularities may first appear as patterns in episodic traces, but they cease to be memory proper once the harness promotes them into explicit reusable guidance. At that point they belong to the skill layer rather than the memory layer.

Taken together, these layers show that memory externalizes not a single homogeneous database but the temporal burden of continuity at multiple levels of abstraction. Working context supports immediate resumption, episodic records support reflection and recovery, semantic memory supports abstraction and transfer, and personalized memory supports cross-session adaptation to users and environments. A harness must treat these stores differently because each one changes a different part of what the model would otherwise have to recover internally.

\FloatBarrier

\subsection{How It Is Externalized: Memory Architectures}
\label{memory: how}

When these layers are externalized, the main design question becomes how aggressively active reasoning is separated from stored state. Following the taxonomy of \citet{du2026memory}, current systems can be read as four broad architectural paradigms: Monolithic Context, Context with Retrieval Storage, Hierarchical Memory and Orchestration, and Adaptive Memory Systems. The progression is not just toward larger stores. It is toward more explicit policies for what gets written, promoted, retrieved, compressed, or forgotten.

\subsubsection{Monolithic Context}
Early systems relied on monolithic context: all relevant history, or a summary of it, remained directly in the prompt. This design is transparent and easy to prototype because no separate memory service is required, and for short tasks it can work surprisingly well. Its limitations are structural. Capacity scales poorly, summaries drift, and the model must spend scarce tokens both carrying history and solving the present step. Most importantly, the state disappears with the session, so the agent does not accumulate durable experience.

\subsubsection{Context with Retrieval Storage}

The dominant next step is to keep only near-term working state in context while storing longer-horizon traces externally and retrieving them on demand. This ``context plus retrieval store'' pattern underlies most practical memory systems in production copilots, assistants, and coding agents. It solves the raw capacity problem, but it turns memory quality into a retrieval problem. If the wrong records are surfaced, the model is distracted; if the right ones are missed, the system behaves as though it never remembered them at all.

Recent work attacks this bottleneck from several directions. GraphRAG~\citep{edge2024local} adds graph structure and community-level retrieval, ENGRAM~\citep{cheng2026conditional} compresses memory into latent state representations, and SYNAPSE~\citep{zheng2023synapse} uses spreading activation over a unified episodic-semantic graph to recover less local forms of relevance. These approaches differ in mechanism, but they share the same goal: replacing flat similarity search with a representation better matched to long-horizon reasoning.

\subsubsection{Hierarchical Memory and Orchestration}
Once flat retrieval proves insufficient, systems move to hierarchical memory and orchestration. The key idea is that not every trace deserves the same retention policy or retrieval path. Frameworks such as Mem0~\citep{chhikara2025mem0}, Memory-R1~\citep{yan2025memory}, and Mem-$\alpha$~\citep{wang2025mem} introduce explicit operations for extraction, consolidation, and forgetting, turning memory into a managed lifecycle rather than a passive store. Two design tendencies dominate this space:

\begin{itemize}
    \item \textbf{Resource decoupling in spatio-temporal dimensions.} One branch borrows the logic of operating systems and treats memory as a constrained resource that must be actively managed. MemGPT~\citep{packer2023memgpt} and MemoryOS~\citep{kang2025memory} separate hot working state from colder long-tail storage and swap information across tiers as task demands change. The gain is higher effective capacity under fixed context budgets.
    \item \textbf{Semantic decoupling in cognitive functional dimensions.} A second branch organizes memory by function or content type so that heterogeneous records are not all routed through the same channel. MemoryBank~\citep{zhong2024memorybank} and MIRIX~\citep{wang2025mirix} separate events, user profiles, and world knowledge; MemOS~\citep{li2025memos} distinguishes explicit and implicit memory; and xMemory~\citep{hu2026beyond} builds a topic-event hierarchy. The goal is not simply neat taxonomy, but more precise retrieval under complex task conditions.
\end{itemize}

\subsubsection{Adaptive Memory Systems}

The architectures above still rely heavily on human-designed heuristics. Adaptive memory systems go further by making modules, routing decisions, or retrieval strategies responsive to experience. Two directions are especially visible:


\begin{itemize}
    \item \textbf{Dynamic modules.} Some systems adapt the architecture itself at runtime. MemEvolve~\citep{zhang2025memevolve} decomposes the memory lifecycle into separate encode, store, retrieve, and manage modules that can evolve independently during execution. MemVerse~\citep{liu2025memverse} maintains a short-term cache and a multimodal knowledge graph while periodically distilling fragmented experience into more abstract knowledge and lightweight neural components.
    \item \textbf{Feedback-based strategy optimization.} Other systems keep the architecture relatively fixed but learn better control policies. MemRL~\citep{zhang2026memrl} updates retrieval behavior through non-parametric reinforcement learning. The adaptive framework proposed by ~\citet{zhang2025learn} uses mixture-of-experts gating to route queries dynamically, and GAM~\citep{yan2025general} refines retrieval conditions over multiple rounds of interaction.
\end{itemize}

Across these stages, the major transition is from storage to control. Monolithic context solves existence, retrieval stores solve capacity, hierarchical systems solve organization, and adaptive systems begin to solve policy. Memory therefore ceases to be a passive appendix to prompting. In mature agents it becomes part of the harness control surface that determines what past the model can effectively act on.

\subsection{Memory Demands of the Harness Eras}
\label{memory: harness}

As agents evolve into the Harness era, memory systems are no longer merely isolated storage modules; instead, they become the substrate through which the runtime coordinates continuity, procedural reuse, and governed interaction. The question is no longer only how to store more information, but how to make temporal state selectively legible to planning, execution, and recovery loops.

The Harness environment therefore requires memory systems to explicitly separate state from context. In tasks with extremely long time horizons, the unrestricted accumulation of session history can cause the model to lose track of its attention mechanism. Frameworks such as InfiAgent~\citep{yu2026infiagentinfinitehorizonframeworkgeneralpurpose} propose a file-centric state abstraction, advocating for the file system as the sole authoritative record of task state, where everything—from high-level planning to intermediate variables and tool outputs—must be written in real time. At each decision step, the agent no longer reads lengthy history but instead reads a curated snapshot of the workspace and a small number of recent actions. This is the harness-level expression of memory's core representational role: not preserving all history in prompt, but materializing the current state in a form the model can act on.

Memory must also be integrated with the skill system, but the two layers play different roles. Memory stores the evidence of prior execution: traces, outcomes, failures, and user- or task-specific context. Skills begin only when some of that evidence is promoted into explicit reusable procedure. In the opposite direction, every skill execution produces new traces that must be written back into memory. Memory is therefore not itself procedural guidance; it is the evidence base from which such guidance can later be derived.

Protocol coupling imposes a further requirement. Tool results, approvals, delegation events, and external state transitions may arrive through protocolized interfaces, but they become memory only once they are normalized and written into persistent state. Conversely, memory retrieval may influence which protocol path should be chosen next. In a mature harness, memory and protocol are linked by a governed read/write loop, but they remain conceptually distinct: protocol governs exchange, while memory governs persistence across time.

Finally, sharing and governance mechanisms become mandatory once multiple agents rely on common externalized state. Establishing read/write permissions for memory, resolving conflicts among stored facts, and controlling each agent's access quota to shared knowledge all require low-level control capabilities comparable to those of an operating system. Memory in the harness era is therefore best understood as managed state infrastructure: it externalizes temporal burden, reshapes what the model must remember internally, and provides the persistent substrate on which the rest of the harness operates.

\subsection{Memory as Cognitive Artifact}
\label{memory: cognitive artifact}

The preceding sections surveyed the content, architecture, and harness integration of memory systems. This final section steps back to interpret what memory externalization achieves as a representational transformation, drawing on Norman's theory of cognitive artifacts~\citep{norman1993things} and Kirsh's account of complementary strategies~\citep{kirsh1995complementary}.

Modern LLMs are stateless generators: each call begins with a fresh context, so continuity must be reconstructed rather than carried forward. In short interactions, that limitation can be hidden inside the prompt. In long-horizon work, it becomes structural. Past attempts, partially completed work, user-specific facts, and environmental state cannot all remain live in context without cost, drift, and eventual truncation. The original task facing a bounded model is therefore intractable in principle: keep an effectively unbounded history available while still reasoning clearly about the present.

Memory externalization changes the structure of that task. In Norman's terms, the representational transformation converts an internal recall problem into an external recognition-and-retrieval problem. The model no longer has to recover relevant history from its parameters; it has to recognize and use a curated slice of history that the memory system has already surfaced. This is closely analogous to Norman's analysis of how an external list changes the nature of remembering: the crucial point is not that extra information has been added, but that the form of the cognitive task itself has been reorganized~\citep{norman1991cognitive}. The same shift was identified in Section~\ref{sec:bg_context} at the context level; memory extends it across sessions and time horizons that no single context window can span.

This interpretation clarifies why retrieval quality matters more than raw storage capacity. A system with vast storage but weak retrieval still presents the model with the wrong problem representation: the history exists, but the task has not been transformed. By contrast, a modest store with strong indexing, summarization, and contextual selection can make downstream reasoning significantly easier. The success criterion for memory is therefore not ``how much did we save?'' but ``did we make the current decision legible?''

The same perspective also illuminates Kirsh's notion of complementary strategies, according to which agents improve performance not only by thinking harder internally but also by reorganizing the external environment so that some cognitive work is offloaded into it~\citep{kirsh1995complementary}. Memory systems implement exactly this strategy for the temporal dimension. Rather than forcing the model to carry all relevant state internally, the harness externalizes persistence, freshness management, and relevance filtering, while leaving interpretation and contextual judgment to the model. The division is complementary: each side handles the part of the task it does best.

The cognitive-artifact view also explains common failure modes as failures of representational design rather than mere implementation bugs. Stale memories misrepresent the present by offering an outdated problem representation. Over-abstracted memories lose the operational details needed for the current decision. Under-abstracted memories flood the prompt with noise, degrading the very recognition task that externalization was supposed to simplify. Poisoned or conflicting memories contaminate future reasoning by embedding incorrect premises into the retrieved slice. In each case, the memory system has failed not because it stored too little or too much, but because it did not transform history into a usable present.

Seen in this light, memory is not simply an engineering convenience for expanding effective context. It is a cognitive artifact that reshapes the temporal burden of agency. By converting unbounded recall into bounded, curated retrieval, it changes the task the model faces at every decision point. That transformation is what connects the architectural progression surveyed in this section---from monolithic context through adaptive systems---to a single underlying design goal: making the right history legible at the right moment, so that the model's fixed inferential capacity is spent on reasoning rather than on remembering.

%% file: text/04_skills.tex
\section{Externalized Expertise: Skills}
\label{sec:skills}
Skill externalization addresses the procedural burden of agency. A language model may know, in principle, how to solve a task, yet reliable execution still requires reconstructing workflows, defaults, and constraints each time a task is attempted. That burden grows with task length, environmental specificity, and the number of branching decisions, and it manifests as variance: omitted steps, unstable tool use, and inconsistent stopping conditions.

The representational shift introduced by skills is therefore from repeated synthesis to reusable procedure. Instead of asking the model to regenerate task-specific know-how from weights or ad hoc prompts on every run, a skill system packages that know-how into explicit artifacts that can be discovered, loaded, revised, and composed. This does not mainly expand the set of actions available to the agent; it changes the task the model faces at runtime from inventing a workflow to selecting and following one~\citep{xu2026agentskillslargelanguage,wang2026skillorchestralearningrouteagents}.

In harnessed agents, skills sit between memory and action. They are often selected in light of retrieved state, bound to tools and subagents through protocolized interfaces, and updated from execution traces and post hoc reflection. As discussed in Section~\ref{sec:memory}, memory externalizes what has been learned over time; skills externalize how that accumulated experience becomes a reusable operating structure~\citep{sumers2024coala,wu2026agent}. The chapter therefore focuses on three linked questions: what burden skills externalize, how skills reorganize task execution, and how they become actionable inside a larger harness.


\begin{figure}[!htbp]
  \centering
  \includegraphics[width=\linewidth]{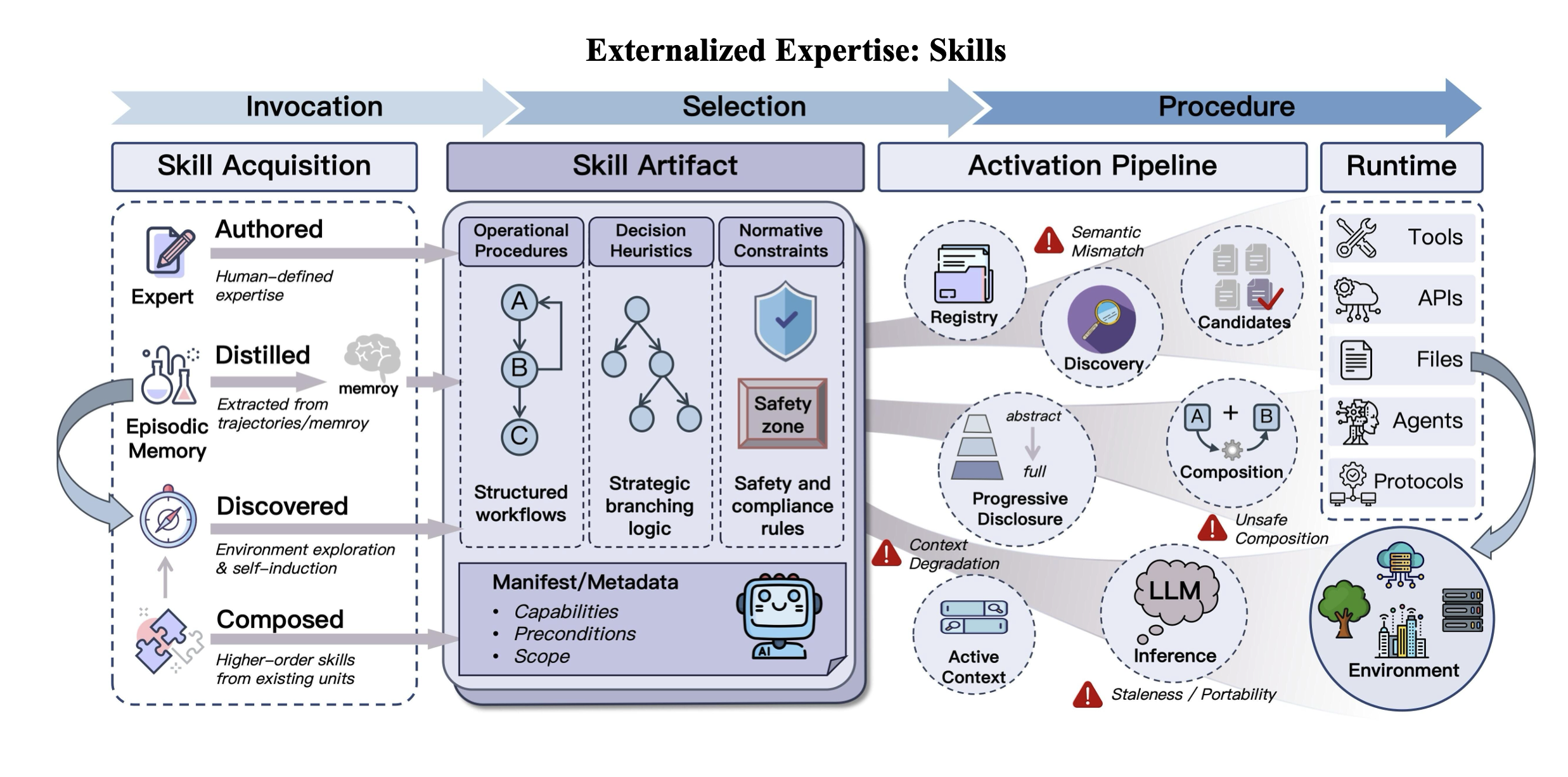}
  \caption{\textbf{Skills as externalized expertise.} The figure traces the full lifecycle of a skill through three phases---invocation, selection, and procedure. \emph{Skill Acquisition} shows four pathways by which procedural know-how enters the system: authored by experts, distilled from episodic memory and trajectories, discovered through environment exploration and self-induction, or composed from existing units. \emph{Skill Artifact} packages that know-how into operational procedures, decision heuristics, and normative constraints, accompanied by a manifest declaring capabilities, preconditions, and scope. \emph{Activation Pipeline} handles registry-based discovery via semantic abstraction, progressive disclosure from abstract summaries to full guides, and composition that binds skills to tools, APIs, files, agents, and protocols. \emph{Runtime} shows how the active context and the LLM execute the selected skill, while boundary conditions---staleness, portability limits, context-dependent degradation, and unsafe composition---constrain reliability.}
\label{fig:skill_externalization}
\end{figure}

\FloatBarrier

\subsection{What is Externalized: Procedural Expertise}
Skill externalization concerns procedural expertise rather than isolated action interfaces. Expertise here means a repeatable way of carrying out a task under recurring assumptions and constraints, not a vague claim that the model ``can'' do something. A useful boundary follows from that definition: tools expose operations, protocols govern how those operations are described and invoked, and skills encode how a class of tasks should be executed with them. In practice, that expertise has three coupled components: \textbf{operational procedures}, \textbf{decision heuristics}, and \textbf{normative constraints}. Together they define the reusable unit of know-how that a harness can externalize.

\subsubsection{Operational Procedure}
Operational procedure is the task skeleton: the decomposition of a complex job into steps, phases, dependencies, and stopping conditions. It addresses a common failure mode in LLM agents. Many errors do not come from incapacity at the action level; they come from instability at the process level, such as skipped steps, misordered operations, or premature termination~\citep{hsiao2025procedural,nandi2026sopbench}. Externalizing procedure turns that fragile process knowledge into an explicit operating path.

This shift has deep roots in the broader evolution of LLM reasoning. Chain-of-Thought made intermediate reasoning explicit~\citep{wei2023chainofthought}; ReAct coupled reasoning with action~\citep{yao2023react}; later prompt-chaining and orchestration systems packaged recurring patterns into engineered workflows. What those approaches often lacked was persistence. The procedure existed in the current run, but not yet as a reusable artifact. Skill systems close that gap by turning workflow structure into something that can be stored, revised, and reapplied~\citep{ye2025sopagent}.

Once procedures are externalized, execution becomes less improvisational. The agent can resume after interruption, hand work across contexts or collaborators, and recover state without reconstructing the entire workflow from memory. This matters most in long-horizon, multi-agent, and production settings, where process stability is often more important than momentary fluency.

\subsubsection{Decision Heuristics}
If procedures define the skeleton of execution, decision heuristics govern what happens at branches. Real tasks rarely unfold as fixed pipelines. Tools fail, observations are noisy, and several locally plausible actions may compete. Under those conditions, good performance depends on practical rules of thumb derived from experience rather than on exhaustive search alone~\citep{gigerenzer2011heuristic}.

Externalizing those heuristics changes the distribution of reasoning effort. Instead of forcing the model to rediscover local policy at every junction, the system can encode default choices, escalation rules, or preference orderings that have already proved useful. That reduces deliberation cost and also makes behavior more stable. Heuristics are therefore not a secondary convenience. They are one of the main ways a skill captures expert style: what to try first, when to back off, what evidence is sufficient, and which trade-offs are preferred when multiple paths remain viable.

\subsubsection{Normative Constraints}
The third component is normative constraint: the conditions under which a procedure counts as acceptable. A workflow may be technically effective and still be noncompliant, unsafe, or operationally wrong. In real deployments, execution is bounded by testing requirements, scope limits, access restrictions, traceability expectations, and domain-specific operating rules~\citep{chen2021evaluatinglargelanguagemodels,bai2022constitutionalaiharmlessnessai,wei2023chainofthought,schick2023toolformer,madaan2023selfrefineiterativerefinementselffeedback}.

Once externalized, those constraints stop being merely post hoc evaluation criteria and become part of the skill itself. They can shape preconditions, block unsafe branches, require intermediate validation, or define evidence that must be produced before completion. This is what lets skills encode not only how to perform a task, but also how to perform it within organizational and safety boundaries. In mature systems, that makes skills carriers of governance as much as carriers of capability.

Taken together, operational procedures provide structure, decision heuristics provide local policy, and normative constraints provide acceptable boundaries. A skill is reusable only when all three are specified well enough to survive across tasks, contexts, and runs. That is why skills sit above action interfaces and beside memory: they externalize not past state and not raw execution primitives, but repeatable task know-how.

\subsection{From Execution Primitives to Capability Packages}
Skill systems do not emerge in isolation, but they should also not be conflated with tool use. Historically, skills are downstream of two earlier developments: reliable action invocation and large-scale action selection. Those stages expanded what an agent could \emph{do}, but not yet how a class of tasks should be carried out repeatedly. Skills appear only when procedural organization itself becomes an explicit reusable artifact.

\subsubsection{Stage 1: Atomic Execution Primitives}
The first stage equips language models with reliable action execution, for example through structured tool invocation and function-calling interfaces. Toolformer is representative in showing that models can learn when to call tools, how to construct arguments, and how to incorporate results \citep{schick2023toolformer}. The key achievement at this stage is stable access to atomic action units. What it does \emph{not} provide is an explicit reusable procedure for completing a broader class of tasks. The unit is the action primitive, not the skill.

\subsubsection{Stage 2: Large-scale Primitive Selection}
As the number of callable tools grows, the problem shifts from invocation to selection. Work such as Gorilla, ToolLLM, ToolNet, ToolScope, and AutoTool shows that models can retrieve, rank, and dynamically choose among large tool collections \citep{patil2023gorilla,qin2023toolllm,liu2024toolnet,liu2025toolscope,zou2025autotool}. This is a major step toward scalable action selection, but the unit remains the tool rather than the procedure. Even when multi-step behavior begins to emerge, the know-how for accomplishing a task class is still largely implicit in prompts or parameters rather than externalized as a bounded reusable artifact.

\subsubsection{Stage 3: Skill as Packaged Expertise}

The third stage marks a further shift in abstraction. The central question is no longer whether a model can invoke a function or retrieve an appropriate API, but whether the know-how required to complete a class of tasks can be packaged into reusable capability units. 
In this stage, the fundamental unit of capability is no longer an isolated tool call, but a higher-level artifact centered on reusable procedural guidance and execution structure \citep{wang2025inducing,chen2026cua}.
Rather than merely specifying what can be done, a skill increasingly captures how a task should be carried out through reusable procedural organization \citep{li2026skillsbench}.

Recent work makes this transition increasingly explicit. 
Program-based skill induction compiles primitive actions into higher-level reusable skills, showing that agent capabilities can be represented as executable procedural abstractions rather than one-off invocations \citep{wang2025inducing}. 
In web environments, interaction trajectories can be distilled into reusable skill libraries or skill APIs, allowing agents to accumulate and refine transferable know-how across tasks \citep{zheng2025skillweaver}. 
In computer-use settings, skills are further organized as parameterized execution and composition graphs, with retrieval, argument instantiation, and failure recovery operating at the skill level rather than the level of individual interface actions \citep{chen2026cua}. 
Related work on SOP-guided agents likewise shows that domain expertise can be externalized as explicit procedural structures that guide execution according to domain-specific procedures \citep{ye2025sopagent}.

Compared with earlier stages, the key transformation here is representational rather than merely operational. 
Capability is no longer treated primarily as access to tools or APIs, but increasingly as packaged procedural knowledge that can be loaded, reused, and composed across tasks \citep{li2026skillsbench,xu2026agentskillslargelanguage}. 
In this sense, Stage 3 does not simply make tool use more complex, but it reflects a shift toward representing agent capability as externalized and reusable procedural know-how.

\subsection{How Skills Are Externalized}

Skill externalization is not exhausted by writing down instructions. 
In mature agent systems, the crucial issue is whether procedural expertise can be represented in a form that is discoverable, loadable, interpretable, bindable, and executable at runtime. 
Therefore, skill externalization involves both a representational layer and a runtime layer. 
The former determines how a skill is described and delimited, while the latter determines whether it can actually function as a reusable capability during task execution~\citep{xu2026agentskillslargelanguage}. In harness terms, a skill only becomes real when the runtime can decide when to load it, which memory to condition it on, and which tools, files, or subagents to bind it to. That binding requirement does not make skills identical to tools or protocols; it simply means that procedural expertise must eventually be grounded in executable interfaces.

\subsubsection{Specification}
The externalization of a skill begins at the specification layer. 
Typical forms include \texttt{SKILL.md}, instruction files, manifests, or other declarative specification artifacts. 
These artifacts describe what a skill does, what scenarios it applies to, what dependencies it assumes, what constraints it must satisfy, and under what input-output conditions it should operate. 
A skill specification resembles API documentation more than API implementation.
Its value lies in turning procedural expertise from an opaque internal state into an explicit object that can be inspected, discussed, revised, and governed~\citep{ling2026agent}.

A well-formed skill specification should ideally cover at least five kinds of information, namely capability boundaries, scope of applicability, preconditions, execution constraints, and examples together with counterexamples. 
The first two clarify what kinds of problems the skill is intended to solve. 
The next two clarify when it can be used safely and under what operating assumptions. 
The final category helps anchor the intended usage pattern in concrete cases, thereby reducing underspecified interpretation by the model.
Through such structured specification, a skill is elevated from an unstructured prompting trick to a bounded capability description, which in turn provides the foundation for discovery, loading, version control, and governance.

\subsubsection{Discovery}
Once skills become explicit artifacts, they naturally introduce the problem of registration and discovery. 
In realistic settings, an agent cannot indiscriminately load every available skill for every task. 
It therefore requires some form of registry and discovery mechanism to support selective retrieval. 
A skill may be published to a local repository, an organizational registry, or a platform-level marketplace, while the agent searches for relevant candidates based on task goals, context state, and environmental conditions \citep{zheng2025skillweaver}.

This discovery process may rely on semantic retrieval, structured metadata, task decomposition, or combinations of these strategies, depending on the system design.
The key point is that the system is not merely asking which tool can be called. 
It is asking which unit of procedural expertise is appropriate for the present problem. 
This makes skill discovery a higher-level matching problem.
It must consider not only topic similarity, but also task complexity, environmental assumptions, operational constraints, and risk conditions.
A skill should therefore be retrieved not simply because its keywords overlap with the task description, but because it is genuinely compatible with the semantic and operational structure of the current task \citep{ross2025when2call}. 
Skill externalization is incomplete if a skill is merely stored. 
It must also be retrievable under realistic task conditions.

\subsubsection{Progressive Disclosure}
The discovery of a skill does not imply that its full contents should immediately be injected into the active context. 
Because long context does not reliably translate into better performance, detailed instructions can become a source of reasoning noise rather than a source of guidance. 
For this reason, current skill systems often benefit from a progressive disclosure strategy in which the existence of a skill is exposed first, and deeper detail is loaded only when needed \citep{xu2026agentskillslargelanguage}.

In current industrial implementations, this often takes a layered form.
At a minimal level, the model sees only the name of the skill together with a brief description, which is sufficient to signal that the capability exists. 
A deeper level may expose manifest-like information such as applicability conditions, required prerequisites, and major constraints.
Only at the deepest level does the system load the full guide, including detailed procedures, exception handling, examples, and supporting files.
The purpose of such staged loading is not simply to compress documentation. 
More fundamentally, it turns the question of whether more skill detail is needed into a runtime decision in its own right.
In this way, the informational density of the skill can be matched to the complexity of the current task rather than saturating the context with unnecessary detail from the outset.
This design is especially visible in current industrial implementations of skills, such as Claude Code's skill system~\citep{anthropic2025skills}.

\subsubsection{Execution Binding}
A skill remains a cognitive-level description unless it is connected to executable action. Actual task completion therefore depends on a binding process that translates the natural-language or structured specification of a skill into concrete operations in the current environment. It is precisely at this point that the distinction between skills, tools, and protocols becomes clear.

A skill is usually not itself an action executor. Instead, it must be bound to a lower-level runtime substrate, such as tools, files, APIs, sub-agents, protocol endpoints, or other execution interfaces. A skill may specify that the agent should search relevant code, run tests, and summarize the resulting diff, but the actions themselves are carried out by search tools, file operations, shell commands, and test runners. Tools therefore provide the executable operations; protocols govern how those operations are described and invoked; skills provide the higher-level strategy for combining them into repeatable task completion.

This binding typically requires an intermediate interpretation layer that determines, in the current context, which skill steps should be activated, which primitives should be bound, which conditions should trigger branching, and which constraints should take priority. 
Without such an interpretation and binding process, a skill easily remains a static artifact that is readable in principle but unusable in practice. 
More generally, schema-based interfaces such as MCP~\citep{anthropic2024mcp} support this runtime binding layer by making capabilities discoverable and invocable without collapsing skills into tools or protocols themselves.

\subsubsection{Composition}
The value of a skill system is most fully realized when skills can be composed. 
Unlike atomic tools, skills can participate in higher-order structured coordination, allowing complex tasks to be decomposed into the cooperation of multiple capability packages. 
Common composition patterns include serial execution, parallel division of labor, conditional routing, and recursive invocation of sub-skills within a higher-level skill~\citep{wang2023voyager}.

This compositionality means that a skill is not merely a document intended for model consumption, but a schedulable runtime unit inside an agent architecture. 
More importantly, composition is not just the concatenation of multiple procedural fragments. 
It is a higher-level reuse of procedural expertise itself. 
For example, a skill for producing a data analysis report need not be implemented as a monolithic end-to-end procedure. 
It can instead be organized as a coordinated composition of smaller skills for data cleaning, statistical analysis, visualization, and narrative synthesis.
In this way, the system gains not only stronger task performance, but also better maintainability, replaceability, and auditability.
Composition therefore marks the point at which skills become a genuine capability layer rather than a collection of isolated recipes~\citep{yu2025polyskill}.

Overall, skill externalization should not be understood as the mere publication of a static instruction file. 
It is a coordinated process in which procedural expertise is specified, made discoverable, selectively disclosed, bound to executable substrates, and composed into larger capability structures. 
What matters is not only whether a skill can be written down, but whether it can reliably enter the agent's runtime as a usable unit of action that interoperates with retrieved state and protocolized interfaces.
Hence, the externalization of skills marks a shift from informal prompting toward a more explicit capability layer for agent systems.

\subsection{Skill Acquisition and Evolution}

A skill system matters not only because it stores authored instructions, but because it provides a pathway for turning successful behavior into reusable expertise. 
Skill acquisition is therefore better understood as an evolutionary process in which procedural knowledge is written, extracted, discovered, and recomposed over time~\citep{xu2026agentskillslargelanguage}.

\paragraph{Authored.}
Manual authoring remains the most common and stable route by which skills enter current systems.
Whether in the form of \texttt{SKILL.md}, \texttt{AGENTS.md}, project-level instruction files, or organizational SOP templates, these artifacts are all instances of human-designed procedural capability packages.
Their importance lies not only in providing initial capability, but also in supporting iterative revision. 
When an agent repeatedly exhibits a failure pattern in deployment, engineers can update the corresponding skill so that one observed failure becomes a clarified procedure or an added constraint. 
In this way, authored skill documentation is not merely descriptive. 
It also serves as a practical interface through which operational experience is gradually turned into reusable behavioral structure \citep{ling2026agent}.

\paragraph{Distilled.}
Skills may also be induced from historical trajectories, practice traces, or other stored experience.
Episodic records preserve what the agent previously did and why a trajectory succeeded or failed. 
When certain successful structures recur across tasks, the system can abstract these patterns into more stable procedural units. 
In this sense, memory preserves experience, while skill induction extracts the reusable structure within it. 
Existing evidence supports this most directly when the process is framed as induction from interaction traces rather than as a broad claim that memory automatically becomes skill. 
Skill Set Optimization, for instance, extracts transferable skills from rewarding sub-trajectories \citep{nottingham2024skill}. 
In memory-management settings, MemSkill further shows that some memory operations themselves can be reformulated as learnable and evolvable skills \citep{zhang2026memskill}.

\paragraph{Discovered.}
Beyond manual authoring and post hoc distillation, agents may also autonomously discover new skills through environmental interaction.
Voyager provides an influential example in the Minecraft setting, where exploration, execution feedback, self-verification, and curriculum-driven task selection jointly produce an ever-growing skill library of executable code \citep{wang2023voyager}.
More recent work suggests that this discovery process can also be oriented toward generalization. 
PolySkill, for example, improves skill reuse by separating abstract goals from concrete implementations \citep{yu2025polyskill}. 
Once an agent can identify behavioral patterns that repeatedly succeed and elevate them into explicit skills, the skill library becomes not only a storage layer but also a mechanism for capability growth.

\paragraph{Composed.}
Finally, skills can evolve through composition. 
Many higher-level capabilities are not invented from scratch, but assembled from existing lower-level or mid-level skills. 
A complex workflow such as report generation or code repair may emerge from the repeated coordination of smaller capabilities. 
Composition matters here not only as an execution strategy but also as an acquisition mechanism. 
Once a particular combination of existing skills is repeatedly validated as effective, that combination can itself be packaged as a new higher-level skill. 
In this way, composition generates new reusable units and gradually gives rise to hierarchical skill repertoires rather than flat lists of isolated capabilities \citep{wang2025inducing}.

Overall, skill acquisition is not a one-time design step but a continuing process of writing, extracting, discovering, and recomposing procedural knowledge. 
A mature skill system is therefore defined less by how many instructions it stores than by how effectively it turns experience into reusable externalized expertise. In a harnessed agent, this evolutionary loop is itself systematized: memory provides the evidence, evaluators decide what merits promotion, and protocolized execution surfaces determine whether a candidate skill can actually be deployed.

\subsection{Boundary Conditions}

Skill externalization improves reuse and governance, but it does not guarantee reliability. 
Once procedural expertise is externalized as an explicit artifact, its effectiveness becomes conditional on how well the artifact matches the task, the environment, and the runtime in which it is used.
In practice, the main boundary conditions concern semantic alignment, portability and staleness, unsafe composition, and context-dependent degradation.

\paragraph{Semantic alignment.}
A skill specification expresses intent and guidance in natural language or lightweight structured form, while actual execution depends on concrete tools, APIs, and environmental constraints. 
As a result, a model may follow the literal wording of a skill while still missing the real objective of the task. 
Existing evidence suggests that the effectiveness of skills depends heavily on the alignment between task intent, skill description, and invocation decision.
SkillProbe identifies semantic-behavioral inconsistency as a fundamental flaw in existing skill marketplaces~\citep{guo2026skillprobe}. 
Related work on tool-use decision making likewise shows that the key difficulty is often not only whether an external capability can be called, but whether it should be called under the current interpretation of the task~\citep{ross2025when2call}.
This suggests that externalized skills remain sensitive to mismatches between description and use.

\paragraph{Portability and staleness.}
Even when a skill is internally coherent, its validity across environments cannot be assumed. 
Changes in websites, APIs, dependencies, workflows, or runtime conventions can make a once-effective skill partially misleading or entirely obsolete. 
More broadly, heterogeneity across agent frameworks, tool substrates, and base models means that the same skill may not behave consistently across settings. 
Programmatic-skill work already shows that some induced skills transfer across websites while incompatible ones must be updated to accommodate environmental change~\citep{wang2025inducing}. 
SkillsBench further indicates that skill utility varies substantially across domains and model-agent configurations~\citep{li2026skillsbench}. 
The broader implication is that skill portability is best treated as a conditional empirical property rather than an intrinsic feature of externalization.

\paragraph{Unsafe composition.}
Composition makes skills more powerful, but it also creates new risks. 
Skills that appear harmless in isolation may interact unsafely when combined, especially when they bundle long-form instructions, executable scripts, and external dependencies.
In such cases, the problem is not confined to a single skill artifact, but emerges from the interaction among multiple artifacts and the interfaces that connect them. 
This is one of the boundary conditions for which direct evidence is now available. 
Large-scale empirical studies of public skill ecosystems report substantial rates of vulnerabilities, including prompt injection, data exfiltration, privilege escalation, and supply-chain risk~\citep{liu2026agent}. 
Attack-oriented studies further show that skill files themselves can become realistic prompt-injection surfaces for current agents~\citep{wang2026skills}. 
Skill composition should therefore be treated as a security-sensitive process rather than a purely benign form of modular reuse.

\paragraph{Context-dependent degradation.}
A further difficulty is that skill execution can degrade over extended interaction. 
Even when a skill file has been updated, the agent may continue to follow outdated operational logic because of residual session context, cached summaries, or previously reinforced action patterns. 
At the same time, detailed skill guides can interfere with global task tracking when too much local procedural detail is injected into the context. 
In such cases, the model may execute the instructions carefully while losing sight of the true success condition. 
Direct skill-specific evidence for these effects is still limited, but adjacent work on multi-turn drift, long-horizon reliability, and long-context reasoning strongly suggests that they are realistic boundary conditions~\citep{lee2026capable}. 
Skill loading should therefore be treated not only as a retrieval problem, but also as a problem of context allocation and execution stability.

Taken together, these boundary conditions show that a skill is not a self-sufficient module that remains stable once written. 
Its effectiveness depends on continued alignment with tasks, environments, runtime conditions, and security constraints.
Skills should therefore be treated not as isolated artifacts, but as components embedded in a broader engineering framework. 
This is precisely why skill design ultimately points beyond the artifact itself toward harness engineering.

\subsection{Skills in the Harness}
\label{skills: harness}

The boundary conditions above show that skills cannot be evaluated as standalone artifacts. Their reliability depends on how they are situated within a running system. This section examines how skills become operational once embedded in a harness, focusing on the couplings that connect them to memory, protocols, and runtime governance.

\paragraph{Conditioning on memory.}
A skill is selected and parameterized in light of retrieved state. The harness queries memory for task history, prior outcomes, user-specific context, and environmental constraints, then uses that evidence to decide which skill to load, which parameters to instantiate, and which branches to prefer. Without this conditioning loop, skill selection degenerates into keyword matching against task descriptions. With it, the same skill can be applied differently depending on what the agent has previously learned. Memory therefore supplies the situational evidence that makes skill choice contextual rather than generic.

\paragraph{Binding through protocols.}
Once selected, a skill must be grounded in executable action. That grounding passes through protocolized interfaces: tool schemas, subagent delegation contracts, file operations, and approval workflows. The harness mediates this binding by resolving which protocol endpoints are currently available, checking permissions, and routing skill steps to the appropriate execution substrates. Skills and protocols are therefore complementary: skills specify what should be done; protocols specify how the resulting actions are described, invoked, and governed.

\paragraph{Runtime governance.}
In production settings, the harness also imposes governance over skill execution. This includes permission checks before sensitive operations, approval gates for high-risk steps, audit logging of which skill was loaded and what actions it produced, and rollback mechanisms when execution fails partway through a multi-step procedure. These controls are not part of the skill artifact itself; they are properties of the harness environment in which the skill runs. A skill that is safe and effective in a sandboxed development context may require additional constraints in a production deployment, and the harness is the layer that enforces those constraints.

\paragraph{Lifecycle feedback.}
Finally, the harness closes the loop between skill execution and skill evolution. Execution traces, success rates, failure patterns, and user corrections are written back into memory. Over time, that evidence may trigger skill revision, deprecation, or the promotion of new candidate skills. The harness therefore does not merely host skills; it provides the feedback infrastructure through which skills improve. This loop connects skill acquisition (Section~4.4) to runtime operation: authored or discovered skills enter the harness, the harness governs their execution, and execution outcomes feed back into the evidence base from which future skills are derived.

\subsection{Skill as Cognitive Artifact}
\label{skills: cognitive artifact}

The following interpretation is primarily theoretical rather than directly empirical.
It draws on classic work on cognitive artifacts to help explain why externalized skills can improve the organization of procedural expertise, rather than to claim that these theories were originally developed for LLM agents.

From the perspective of Norman's theory of cognitive artifacts, a skill system can be understood as a representational transformation along the dimension of capability organization \citep{norman1993things}.
Without externalized skills, a model must repeatedly reconstruct procedural knowledge from internal parameters during task execution. 
With skills, part of that procedural burden is moved into an explicit external representation that can be loaded, inspected, and followed. 
This shifts the task from unstable latent procedural recall toward a more stable process of recognizing applicable guidance and acting under it. 
In that respect, the role of a skill file is closely analogous to Norman's analysis of how an external list changes the nature of remembering. 
The crucial point is not simply that extra information has been added. 
It is that the form of the cognitive task itself has been reorganized.

This reorganization matters because it changes what the model must do at inference time. 
In the absence of a skill, the model must probabilistically recover an appropriate way of proceeding from its parameters under the pressure of the current context. 
Once the skill has been externalized, the procedural structure is already present as an object in the environment. 
The model's burden shifts toward interpreting the current situation, recognizing whether the skill applies, following the relevant guidance, and handling local exceptions. 
Procedural knowledge is therefore no longer something that must be reconstructed from scratch on each run. 
It becomes an external object that can be operated on directly \citep{li2026skillsbench,xu2026agentskillslargelanguage}.

This interpretation also aligns with Kirsh's notion of complementary strategies, according to which agents improve performance not only by thinking harder internally, but also by reorganizing the external environment so that some cognitive work is offloaded into it \citep{kirsh1995complementary}. 
LLMs are often not especially reliable at reproducing long multi-step procedures in a stable and repeatable manner. 
The same prompt may yield different decompositions, branching decisions, or stopping conditions across runs. 
By contrast, they are comparatively better at reading explicit guidance, matching it to the current context, and adapting execution locally under stated constraints. 
A skill can therefore be understood as an engineered complementary strategy.
It externalizes procedure definitions, constraints, and portions of best practice into an artifact, while leaving interpretation, contextual matching, and exception handling to the model itself.

A skill does not simply add more information to the system. 
It changes how capability is organized.
Procedural expertise is moved out of an opaque and difficult-to-audit parameter space into an inspectable, revisable, and composable external structure. 
That is why the significance of skills lies not merely in engineering convenience, but in a deeper reallocation of where know-how resides and how it becomes available for reuse.
Seen in this light, skills are better understood not simply as prompts or tool wrappers, but as cognitive artifacts for organizing procedural competence in agent systems. At system scale, they externalize procedural burden by converting repeated workflow invention into selection, loading, and composition under runtime control.

%% file: text/05_protocols.tex
\section{Externalized Interaction: Protocols}
\label{sec:protocols}

Protocols externalize the interaction burden of agency. A bare model may infer that a tool should be called, a subagent should be delegated to, or a response should be shown to a user, but without explicit contracts it must also improvise message formats, argument structure, lifecycle semantics, permissions, and recovery behavior. That burden turns every external action into a fragile prompt-following exercise.

Within a harness, this protocol layer is where interaction becomes governable. It mediates how tools are discovered, how subagents are contacted, how user-facing state is exposed, how session progress is represented, and how permissions and failures are enforced. A protocol is therefore not a memory store and not a skill description: it specifies the contract by which state, requests, and actions move across system boundaries. The present section therefore examines what interaction burdens protocols externalize, why that externalization matters, how the current protocol landscape is organized, how protocols become operational inside a harness, and how the resulting transformation can be understood through the lens of cognitive artifacts. Section~\ref{protocols: what} identifies the content of interaction that is externalized; Section~\ref{protocols: why} motivates the benefits; Section~\ref{protocols: survey} surveys the protocol families; Section~\ref{protocols: harness} examines harness-level integration; and Section~\ref{protocols: cognitive artifact} closes the chapter with a cognitive-artifact interpretation.

\begin{figure}[!htbp]
  \centering
  \includegraphics[width=0.85\textwidth]{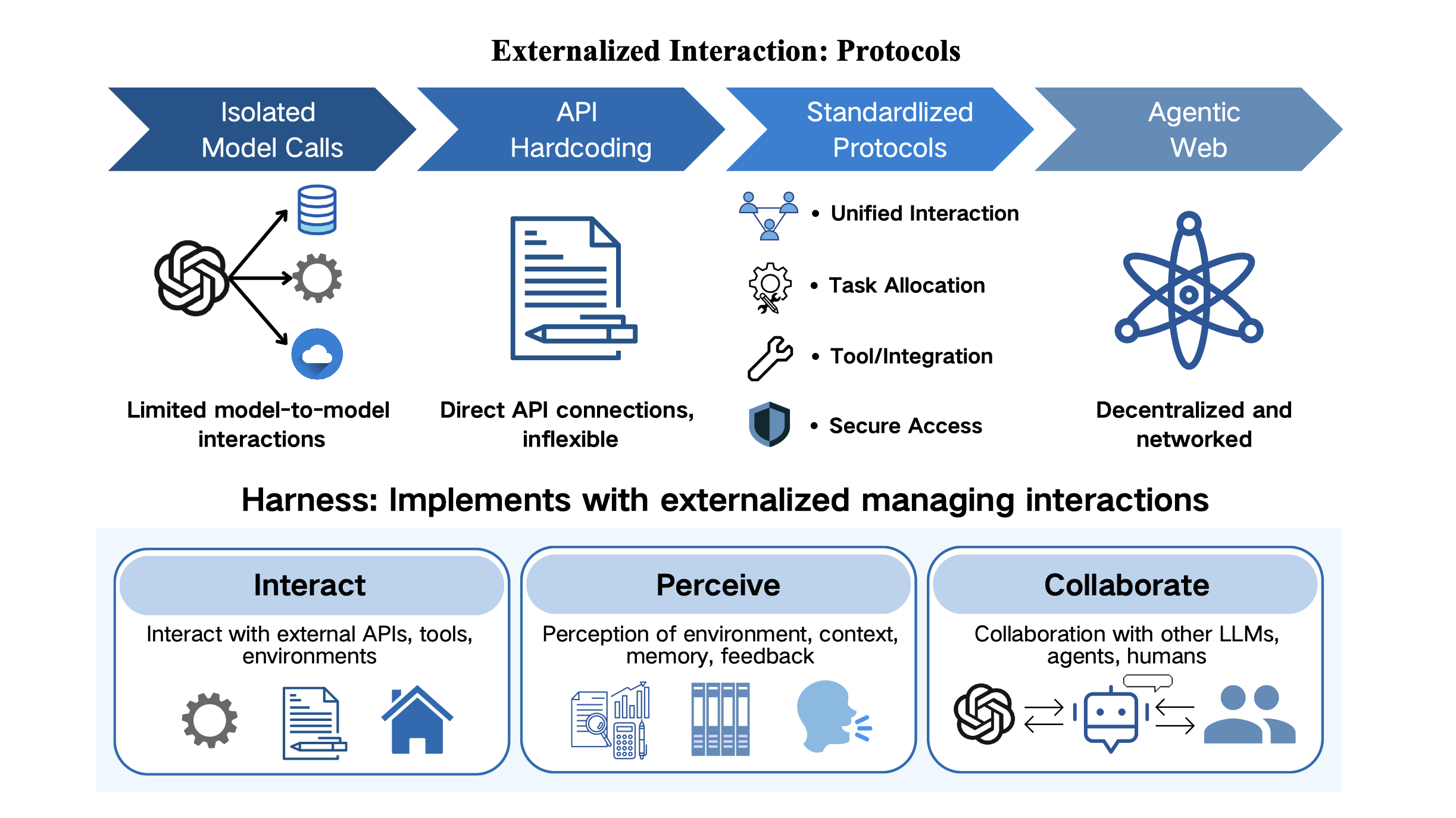}
  \caption{\textbf{Protocols as externalized interaction.} \emph{Upper panel:} The evolutionary trajectory of agent interaction---from isolated model calls with limited model-to-model communication, through hardcoded API connections, to standardized protocols that provide unified interaction, task allocation, tool integration, and secure access, and ultimately toward a decentralized and networked agentic web. \emph{Lower panel:} The harness implements externalized interaction management through three functional surfaces: \textsc{Interact} (interfacing with external APIs, tools, and environments), \textsc{Perceive} (perception of environment, context, memory, and feedback), and \textsc{Collaborate} (collaboration with other LLMs, agents, and humans).}
  \label{fig:protocol_externalization}
\end{figure}
\label{protocols: what}

If memory externalizes temporal state and skills externalize procedural expertise, protocols externalize the contracts that govern how an agent exchanges information and actions with entities outside itself. The representational shift is from free-form communicative inference to structured exchange. Instead of asking the model to invent the syntax and semantics of interaction at runtime, protocols provide typed surfaces, state transitions, and machine-readable constraints that the model can fill and follow. In that sense, protocols do not merely accelerate communication; they change the task from negotiating ad hoc interfaces to operating within explicit contracts.

More concretely, what protocols externalize can be organized along four dimensions:

\paragraph{Invocation grammar.} Every tool call, API request, or delegation message requires a format: argument names, types, ordering, and return structure. Without protocols, the model must infer or reinvent this grammar on each call. Protocols externalize it into schemas and typed interfaces, so the model fills fields rather than guessing syntax.

\paragraph{Lifecycle semantics.} Multi-step interactions need coordination: who acts next, what state transitions are allowed, when a task is complete or has failed. Protocols externalize these sequencing rules into explicit state machines or event streams, removing them from the model's inferential burden.

\paragraph{Permission and trust boundaries.} Real-world agent actions are bounded by who is authorized, what data may flow where, and what evidence must be produced. Protocols externalize these constraints into inspectable rules that a runtime can enforce, rather than relying on the model to self-police.

\paragraph{Discovery metadata.} Before an agent can interact with a tool or another agent, it must know what capabilities are available and how to reach them. Protocols externalize this discovery problem into registries, capability cards, and schema endpoints, replacing implicit prompt-embedded knowledge with queryable metadata.

These four dimensions are not independent---a single protocol may address several at once---but they clarify the scope of what is being externalized. Tools expose operations; skills encode how classes of tasks should be carried out with those operations; protocols specify the interaction grammar, lifecycle, permissions, and discovery mechanisms through which operations and skills become executable across system boundaries.

\subsection{Why Protocols Matter}
\label{protocols: why}
The importance of Agent Protocols follows directly from the burden they externalize: without them, every interaction is partly an inference problem about format, legitimacy, and coordination. Their benefits are easiest to see along three dimensions.

\paragraph{Unified interaction standards.} Protocols give tools, agents, and frontends a shared grammar for discovery, invocation, handoff, and state exchange. Without that layer, the ecosystem fractures into local prompt-plus-parser integrations that do not travel well across runtimes~\citep{yang2025surveyaiagentprotocols}. Standardized interaction makes interoperability a designed property rather than a fortunate accident~\citep{ehtesham2025survey}. It is also the precondition for stable multi-agent collaboration, because delegation and context transfer need common representations before they can be automated.

\paragraph{Improved security, governance, and auditability.} Once agents operate in real environments, the question is not only whether they can act, but whether those actions remain bounded, inspectable, and recoverable~\citep{10.1145/3759355.3759356}. Protocols help by making permissions, identity, execution traces, failure states, and responsibility boundaries explicit. That turns previously implicit glue logic into something a runtime can validate and an operator can audit.

\paragraph{Reduced vendor dependence.} Open interaction contracts also preserve architectural flexibility. If the system accumulates capability at the protocol layer rather than inside provider-specific interfaces, models, vendors, and runtime components can be swapped with less rewiring. Protocols are therefore not only engineering conveniences; they are part of the mechanism by which an agent ecosystem remains portable and evolvable over time~\citep{yang2025surveyaiagentprotocols}.

\subsection{Agent Protocol Survey}
\label{protocols: survey}

In this section, we classify popular Agent Protocols in the community into agent-tool, agent-agent, agent-user, and other protocol families according to the different entities they are designed to interact with, and briefly introduce several representative and commonly used protocols in each category. The purpose of this survey is not to catalogue every emerging standard, but to show that contemporary protocols externalize different slices of interaction burden: some stabilize tool invocation, some stabilize delegation among agents, some stabilize the agent-user boundary, and some govern high-risk vertical workflows.

\subsubsection{Agent-Tool Protocols}
Agent-Tool Protocols were among the earliest protocol families to mature because tool access is where interface fragmentation appears first. MCP~\citep{anthropic2025} is the clearest representative. It provides a standardized way for agents to discover tools, inspect their schemas, and invoke them across heterogeneous services. The problem it addresses is straightforward: without a shared contract, every new tool requires bespoke integration logic, duplicated schema definitions, and provider-specific adaptation.

The boundary with neighboring layers is important. MCP and related protocols specify how tools are described and invoked; they do not specify which multi-step procedure should be followed with those tools, and they do not themselves preserve cross-session cognition once results have been produced. Those roles belong to skills and memory respectively.

Architecturally, MCP turns tool access into protocol-based integration rather than interface-by-interface engineering. Servers expose tools and context resources through a common structure, typically over JSON-RPC 2.0, while clients perform discovery and invocation against that shared specification. This decouples tool ecosystems from model-provider-specific function-calling formats and lowers the cost of adding new capabilities. The practical gains are straightforward: dynamic capability discovery, standardized access to complex external systems, structured request/response exchange, and modular extensibility.

The same separation also improves governance. Because invocation is mediated by a protocol layer rather than emitted as an unconstrained model-generated call, sensitive data handling, permission checks, and audit boundaries can be managed more explicitly. ToolUniverse and related systems extend this logic with more specialized tool schemas and interaction conventions~\citep{gao2025democratizingaiscientistsusing, gao2025txagent}. The broad point is that agent-tool protocols externalize invocation grammar so that tool use becomes portable, inspectable, and scalable rather than an accumulation of bespoke adapters.

\subsubsection{Agent-Agent Protocols}

As soon as multiple agents collaborate, interaction itself becomes a systems problem. Agent-Agent protocols define how capabilities are discovered, how tasks are delegated, how progress and partial state are exchanged, and how results return to the caller. They externalize coordination that would otherwise be buried in prompt conventions or framework-specific glue.

A2A~\citep{google_a2a_2025} is the most visible current example. It standardizes capability discovery through artifacts such as Agent Cards and supports task-oriented communication, state updates, negotiation, and streaming progress between heterogeneous agents. Its importance is not only that agents can message one another, but that delegation becomes structured: the caller can discover what another agent offers, hand off work under a known contract, and track execution without relying on hard-coded assumptions.

Other protocols make different trade-offs. ACP~\citep{ibm_acp_2025} emphasizes lightweight adoption through familiar REST/HTTP patterns and fits settings where compatibility with existing services matters more than rich negotiation. ANP~\citep{chang2025agentnetworkprotocoltechnical} pushes in the opposite direction, aiming at open, Internet-scale interoperability with decentralized identity, cross-domain discovery, and secure end-to-end communication.

Taken together, these protocols show that multi-agent systems need more than message transport. They need standardized semantics for delegation, identity, status, and handoff. That is what lets coordination scale from local orchestration to open agent ecosystems~\citep{yang2025surveyaiagentprotocols,ehtesham2025surveyagentinteroperabilityprotocols}.

\subsubsection{Agent-User Protocols}

Agent-User Protocols formalize the boundary between agent runtimes and user-facing systems. They address a different problem from tool or agent-agent protocols: not how an action is executed elsewhere, but how execution state, outputs, and interface structure are exposed to humans in a form that frontends can render and users can understand~\citep{google_a2ui_github_2025,AGUI}.

A2UI~\citep{google_a2ui_github_2025} represents the interface-generation branch. It lets an agent describe UI structure in a constrained declarative format that host applications can render safely across platforms. The protocol matters because it treats interface construction itself as governed output rather than arbitrary HTML-like text.

AG-UI~\citep{AGUI} represents the streaming-state branch. It standardizes typed execution events such as run start, text emission, tool call arguments, tool call results, completion, and error. Frontends can subscribe to that event stream and render runtime status without learning each framework's private event format.

These two directions are complementary. A2UI externalizes interface composition; AG-UI externalizes the live state transitions behind that interface. Together they show how protocolization makes human-agent interaction more observable, reusable, and portable across hosts.

\subsubsection{Other Protocols}

Beyond general interaction families, some protocols target high-risk vertical workflows where generic interfaces are not enough. UCP~\citep{google_ucp_2026} does this for agentic commerce by standardizing catalogs, requests, and checkout flows so that agents, merchants, and payment providers can interoperate without bespoke integration for every store. AP2~\citep{google_ap2_2025} does the same for payments, emphasizing authorization, signatures, auditability, and proof-bearing transaction objects such as IntentMandate, PaymentMandate, and PaymentReceipt.

These domain protocols matter because they externalize workflow-specific governance, not just generic communication. In vertical settings such as shopping, payments, identity, or compliance, the protocol must encode who is authorized, what evidence must be produced, and how responsibility is tracked across the flow~\citep{ucp_ap2_relation_2026}. Across all families, the common pattern is that protocols make a coordination problem explicit. Tool protocols externalize invocation grammar, agent-agent protocols externalize delegation, agent-user protocols externalize presentation and state streaming, and domain protocols externalize specialized governance.





\subsection{Agent Protocol in Harness Engineering}
\label{protocols: harness}

If the survey above shows which interaction burdens are being externalized in the ecosystem, Harness Engineering shows how those protocol surfaces become part of a running agent. The question is no longer only how an agent ought to communicate with other entities, but how those communication contracts govern execution, persistence, delegation, and recovery once the agent is embedded in a runtime.

Traditional LLM pipelines rely on the model to infer formats, remember recent interaction state, and guess how external actions should be formed. That can be adequate for short, loosely coupled requests, but it breaks down when work spans many steps, tools, agents, or approval boundaries. Harness Engineering externalizes that burden into protocol surfaces. Model outputs are captured as structured intents, validated against permissions and lifecycle state, routed through typed interfaces, and reflected back into the runtime as governed events rather than free-form guesses.


\subsubsection{Intent Capture and Normalization}
Intent capture and normalization is the first of those surfaces. The job of this layer is to translate model-produced language into explicit commands or events that the runtime can validate and act on. Without it, execution semantics remain implicit: the system guesses what the model meant, and small linguistic variations can produce large operational differences.

A mature harness therefore normalizes intent before execution. Free-text proposals are mapped into protocol objects, checked against current context and permission boundaries, and rejected or revised if they do not satisfy the contract. This does not remove model judgment; it relocates the fragile part of the interaction from latent inference to an inspectable interface. The result is higher reliability in long-horizon execution, stronger governance, and cleaner handoffs across tools, agents, and users.

\subsubsection{Capability Discovery and Tool Description}

Capability discovery and tool description form the second surface. In older systems, knowledge of available tools often lives partly in prompts and partly in developer assumptions. Protocolized discovery replaces that with explicit metadata. At session start or phase transitions, the runtime exposes the currently available tools, their schemas, and their input/output structure through standardized messages.

That shift has two effects. It reduces context inflation because the model does not need to carry every tool contract in prompt, and it makes capability boundaries governable because permissions, versioning, and auditing can be enforced against structured metadata rather than inferred from model behavior. In other words, the agent stops guessing what can be called and starts reading a declared capability surface.

\subsubsection{Session and Lifecycle Management}

Harness protocols also need explicit session and lifecycle management because long-horizon agents do not operate as isolated single calls~\cite{chai2025parl}. The runtime must preserve interaction state across multiple turns, context windows, and execution phases. What is preserved here is not durable memory in the full sense, but protocol state: identifiers, roles, pending actions, phase transitions, and allowed next moves.

Most long-running systems therefore treat an execution as a lifecycle object with named states and transition rules. The protocol layer advances that object, emits status changes, and coordinates checkpoint or recovery events. When outputs or checkpoints are written to persistent storage, they become memory. The distinction matters: protocol maintains continuity of interaction; memory maintains continuity across time.

\subsection{Protocol as Cognitive Artifact}
\label{protocols: cognitive artifact}

The preceding sections surveyed the content, landscape, and harness integration of agent protocols. This final section interprets what protocol externalization achieves as a representational transformation, using the same cognitive-artifact framework applied to memory and skills in earlier chapters.

In Norman’s terms, a cognitive artifact transforms a task by changing its representational structure~\citep{norman1993things}. Protocols do this for interaction. Without them, every external action is partly a natural-language inference problem: the model must infer the intended operation, guess the right format, reconstruct acceptable constraints, and hope the receiving system interprets the result correctly. Protocols replace that open-ended inference with a bounded, structured task: fill typed fields, follow a declared state transition, and receive structured feedback. The model still needs judgment about whether and when to act, but it no longer needs to reinvent the syntax and semantics of interaction on each step.

This is one of the strongest forms of externalization in agent systems, because it removes entire classes of reasoning from the critical path. The transformation is analogous to the shift that memory introduces for temporal state (Section~\ref{memory: cognitive artifact}) and that skills introduce for procedural expertise (Section~\ref{skills: cognitive artifact}), but it operates on a different dimension: not what to remember or how to proceed, but how to communicate and coordinate. Standardized protocols reduce the number of decisions that must be made inside the model. They make correct interaction easier and incorrect interaction harder---which is precisely what Norman’s framework predicts when an external representation is well matched to the task.

Kirsh’s account of complementary strategies provides additional clarity~\citep{kirsh1995complementary}. LLMs are strong at interpreting intent, selecting among options, and adapting to context, but they are unreliable at consistently producing well-formed structured output under varying interface requirements. Protocols implement a complementary division of labor: the model contributes judgment and intent, while the protocol surface contributes format, validation, and lifecycle control. Neither side alone is sufficient; together, they produce interaction that is both flexible and disciplined.

This interpretation also explains why protocols serve a distinctive role that cannot be reduced to memory or skills. Memory externalizes what has been learned over time; skills externalize how tasks should be carried out; protocols externalize the discipline by which both memory and skills enter the world as governed action. Memory needs governed read and write paths; skills need bindable interfaces; both depend on protocols to cross system boundaries in a form that is inspectable, auditable, and recoverable. Protocols are therefore not secondary plumbing around a ``real’’ intelligent core. They are cognitive artifacts for interaction---the representational infrastructure that makes other forms of externalized intelligence operational.

%% file: text/06_harness.tex
\section{Unified Externalization: Harness Engineering}
\label{sec:harness}

\begin{figure}[!htbp]
  \centering
  \includegraphics[width=0.85\textwidth]{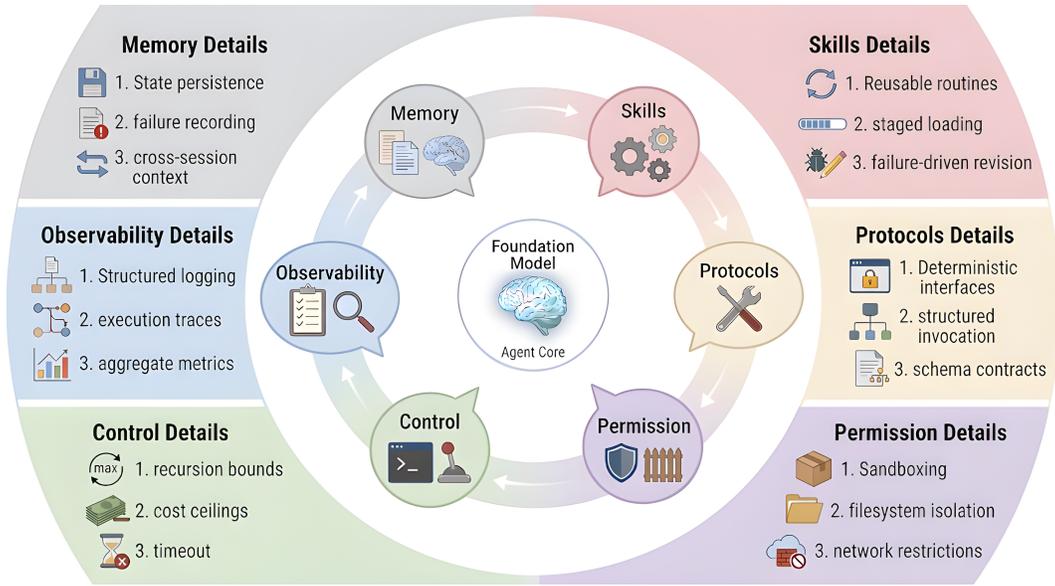}
  \caption{\textbf{The harness as cognitive environment.} The Foundation Model (Agent Core) sits at the center; six harness dimensions form a coordinated ring around it. Three externalization modules---\textbf{Memory} (state persistence, failure recording, cross-session context), \textbf{Skills} (reusable routines, staged loading, failure-driven revision), and \textbf{Protocols} (deterministic interfaces, structured invocation, schema contracts)---supply the externalized cognitive content. Three operational surfaces---\textbf{Permission} (sandboxing, filesystem isolation, network restrictions), \textbf{Control} (recursion bounds, cost ceilings, timeout), and \textbf{Observability} (structured logging, execution traces, aggregate metrics)---govern how that content is accessed, constrained, and monitored at runtime. Arrows indicate the continuous flow among dimensions within the harness loop.}
  \label{fig:harness_environment}
\end{figure}

Figure~\ref{fig:harness_environment} provides an overview: the foundation model sits at the center, surrounded by six harness dimensions that coordinate externalized cognition into coherent agency. Three of those dimensions---Memory, Skills, and Protocols---are the externalization modules analyzed in the preceding chapters (Sections~\ref{sec:memory}--\ref{sec:protocols}). The remaining three---Permission, Control, and Observability---are the operational surfaces that govern how those modules are accessed, constrained, and monitored at runtime. This chapter unpacks these three surfaces into six finer-grained analytical dimensions that together characterize harness design. Each earlier chapter closed by noting that its module becomes fully operational only when embedded in a broader runtime. Sections~\ref{memory: harness}, \ref{skills: harness}, and \ref{protocols: harness} identified specific harness demands from each module's perspective. The present chapter unifies those threads. It asks what kind of system is needed to compose externalized memory, skills, and protocols into coherent agency, and how that system should be understood analytically.

The central claim is that a harness is not merely an implementation convenience layered on top of a capable model. It is the designed cognitive environment within which externalized modules become jointly effective. That framing motivates the structure of this chapter. Section~\ref{harness: what} defines the harness concept and situates it relative to the module-level analyses of earlier chapters. Section~\ref{harness: surfaces} identifies the recurring analytical dimensions along which harness designs vary. Section~\ref{harness: practice} examines how these dimensions manifest in contemporary agent systems. Section~\ref{harness: cognitive environment} closes the chapter by interpreting the harness as a cognitive environment through the lens of distributed cognition and cognitive artifact theory.

\subsection{What is a Harness?}
\label{harness: what}

Externalization, pursued module by module, improves local capability, but agenthood demands global coordination. Memory accumulates experience without specifying which traces are salient to the present task. Skills encapsulate effective routines without automatically incorporating lessons from past interactions. Protocols regularize invocation formats without determining when, or under what policy, a tool should be called. The modules are present, yet the cognitive loop that would render them jointly effective remains under-specified. What is missing is a principled structure that coordinates their interaction over time---aligning perception, memory access, action selection, execution, monitoring, and revision within a single operational envelope.

The term ``harness’’ names that structure. It has recently entered practice as a descriptor for the scaffolding that converts raw model capability into reliable agent behavior. OpenAI’s engineering discussions around Codex, for instance, use the term explicitly to describe the agent loop, execution logic, feedback pathways, and surrounding operational machinery that make the system usable~\citep{openai2025codex}. Because the concept is still consolidating, the characterization we offer here is best understood as a synthesis of recurring patterns in current systems rather than a closed definition.

A practical agent, on this account, is better understood as a model operating inside a harness than as a model with peripheral capabilities attached. A foundation model alone retains general-purpose inference ability, but lacks the operational structure that determines what it can access, how it may act, how its actions are constrained, and how its behavior is observed and revised across time. The harness supplies that structure. It governs the pathways by which the model encounters context, invokes tools, preserves state, and responds to feedback. Agency is therefore not located in the model alone; it emerges from the coupling of the model with the environment that organizes its cognition into action.

Described functionally, the harness comprises the external systems that make such coupling possible: persistent memory and project-level context, reusable skills and executable routines, protocolized interfaces for deterministic interaction with tools and services, and the broader runtime infrastructure within which these elements become operational. The crucial point is not the exact inventory of components---which varies across implementations and will continue to evolve---but their collective role: they create the conditions under which model reasoning can be made stable enough to support sustained work. This shifts the locus of analytical attention from model capability alone to the representational, procedural, and operational conditions under which the model perceives, decides, and acts. Improvements in agency may therefore come not only from better base models, but also from better organization of memory, sharper constraint regimes, more legible feedback channels, and more carefully designed execution environments.

\subsection{Analytical Dimensions of Harness Design}
\label{harness: surfaces}

The modules discussed in earlier chapters---memory stores, skill artifacts, and protocol interfaces---supply the raw materials of externalized cognition, but they do not by themselves specify how the runtime coordinates perception, action, constraint, and feedback over time. That coordination is the province of the harness. The three operational surfaces highlighted in Figure~\ref{fig:harness_environment}---Permission, Control, and Observability---can be decomposed into six recurring dimensions of design variation. Each dimension addresses a distinct aspect of how externalized modules are composed into a functioning agent; together, they provide an analytical framework for comparing harness architectures rather than an implementation checklist.

\subsubsection{Agent Loop and Control Flow}

The agent loop is the temporal backbone of the harness. At its simplest, it implements a perceive--retrieve--plan--act--observe cycle in which the model receives a structured view of the current state, decides on an action, executes it through a tool or protocol interface, observes the result, and updates its internal plan accordingly~\citep{yao2023react,shinn2023reflexion}. Practical systems vary the loop structure considerably. Single-loop designs interleave reasoning and action within one generation pass; hierarchical designs separate a planning agent that decomposes goals from executor agents that carry out individual steps; and multi-agent designs route subtasks across specialized agents with distinct tool sets and permission scopes~\citep{wu2023autogen,hong2023metagpt,langchain2024langgraph}.

What the harness adds beyond a bare loop is governance over termination, recursion, and resource consumption. Without explicit control, an agent loop can cycle indefinitely, escalate costs through unbounded tool calls, or recurse into sub-agent spawns that exhaust context or compute budgets. Production harnesses therefore enforce maximum step counts, recursion depth limits, per-step cost ceilings, and timeout constraints. These controls are not secondary safety measures; they define the operational envelope within which the agent's reasoning unfolds. A well-tuned loop makes the agent more reliable not by making the model smarter, but by bounding the space of possible execution paths.

\subsubsection{Sandboxing and Execution Isolation}

Whenever an agent acts on the world---writing files, executing shell commands, calling external APIs---the harness must decide how much of the environment to expose and how to contain unintended side effects. Sandboxing is the engineering response to that requirement. It creates a controlled execution boundary that limits what the agent can read, write, and modify, and it provides the reproducibility guarantees that make failures diagnosable and rollbacks feasible.

Contemporary systems implement isolation at different granularities. Codex-style agents run each task inside a dedicated cloud sandbox with its own filesystem snapshot, network restrictions, and resource quotas, so that one execution cannot contaminate another~\citep{wang2025openhands,yang2024sweagent}. Claude Code takes a complementary approach by exposing graduated permission modes---from fully autonomous execution to mandatory user approval for every tool call---so that the same agent can operate at different trust levels depending on the task and the operator's risk tolerance~\citep{anthropic2026skillsdocs}. In both cases, the sandbox is not merely a security fence. It is a cognitive boundary that simplifies the agent's operating environment by removing irrelevant state, restricting dangerous actions, and making the workspace inspectable. Isolation thereby serves the same representational function as other forms of externalization: it changes what the model must reason about.

\subsubsection{Human Oversight and Approval Gates}

Full autonomy is rarely appropriate for deployed agents. Most production systems therefore insert intervention points into the agent loop where a human operator can inspect proposed actions, approve or reject them, supply corrections, or redirect execution. The design question is where those gates should be placed and how much autonomy to grant between them.

Three patterns are common. Pre-execution approval pauses the agent before every potentially consequential action and asks for explicit confirmation. Post-execution review lets the agent act but surfaces results for inspection before committing or continuing. Escalation triggers allow the agent to run autonomously under normal conditions but halt and request human input when specific risk signals are detected---such as actions involving sensitive data, irreversible operations, or confidence below a threshold. Hook systems generalize this pattern by allowing operators to attach arbitrary logic---shell scripts, validation checks, notification dispatches---to specific lifecycle events in the agent loop, such as tool invocation, file write, or subagent spawn~\citep{lazaros2026human,fernandez2026agent}. The level of autonomy is therefore not a binary property of the agent but a configurable parameter of the harness, adjustable per task, per tool, and per organizational policy.

\subsubsection{Observability and Structured Feedback}

An agent that acts without leaving inspectable traces is an agent that cannot be debugged, audited, or improved. Observability is the harness surface that makes the agent's internal trajectory visible to developers, operators, and the agent itself~\cite{zhu2026verifiability,zheng2025survey}.

At the implementation level, observability typically involves structured logging of every model invocation, tool call, memory read/write, and decision branch; execution traces that link each action to its causal antecedents; and aggregate metrics such as step counts, token consumption, error rates, and latency distributions. These records serve two distinct purposes. Externally, they support debugging, compliance auditing, and post-incident analysis~\citep{10.1145/3759355.3759356}. Internally, they close the feedback loop that connects execution outcomes back to the modules that produced them. A failed tool call can trigger a memory write that records the failure context; a pattern of repeated failures can flag a skill for revision; a latency spike can cause the harness to switch protocol paths. Without structured observability, these feedback loops cannot operate, and the harness remains a static scaffold rather than an adaptive system. Observability is therefore not an auxiliary convenience; it is the mechanism by which the harness learns from its own operation.

\subsubsection{Configuration, Permissions, and Policy Encoding}

A harness must encode not only what an agent can do, but what it is allowed to do under what conditions. This requires a configuration layer that separates policy from execution logic and makes governance rules explicit, versionable, and auditable.

In practice, configuration is typically stratified across multiple scopes. User-level settings encode personal preferences and trust boundaries. Project-level settings specify which tools are available, which file paths are accessible, and which commands require approval. Organization-level settings impose compliance constraints, cost ceilings, and data-handling rules that individual projects cannot override. This layered model means that the same base agent can operate under different policy regimes depending on its deployment context, without any change to the model or the skill artifacts it loads~\citep{anthropic2026skillsdocs,lee2026structuredapproachsafetycase}. Permissions and policies are therefore best understood as externalized governance: constraints that would otherwise have to be embedded in prompts or enforced through post-hoc filtering are instead encoded as declarative rules that the harness enforces at runtime.

\subsubsection{Context Budget Management}

The context window remains the scarcest shared resource in any agent system. Memory retrieval, skill loading, protocol schemas, tool descriptions, and the model's own reasoning traces all compete for the same finite token budget. How that budget is allocated is a harness-level coordination problem that no single module can solve on its own.

Effective context management typically combines several strategies. Summarization compresses older conversation turns and execution history into shorter representations that preserve decision-relevant information while freeing tokens for the current step~\citep{packer2023memgpt}. Priority-based eviction removes or demotes context entries whose relevance to the active subtask has decayed. Staged loading---already discussed for skills in Section~\ref{sec:skills}---ensures that detailed procedural guidance enters the context only when a matching task pattern is detected, rather than occupying budget from session start. The harness orchestrates these strategies jointly, because the optimal allocation depends on the current phase of execution: an early planning phase may need more memory and less skill detail, while a late execution phase may need the reverse. Context budget management is therefore not a compression problem in isolation. It is a dynamic resource-allocation problem whose solution must be informed by the agent's current goals, the modules it is drawing on, and the constraints under which it operates.

\vspace{1mm}
Taken together, these six dimensions---loop control, sandboxing, human oversight, observability, configuration, and context management---provide an analytical framework for characterizing harness architectures. None of them is a form of externalization in its own right; each is part of the coordinative infrastructure that makes memory, skills, and protocols function as a coherent system. The next subsection uses this framework to examine how contemporary agent systems instantiate these dimensions in practice.

\subsection{Harness in Practice: Contemporary Agent Systems}
\label{harness: practice}

The analytical dimensions identified above are not abstract desiderata; they correspond to concrete design choices observable across deployed agent systems. Contemporary production agents---such as OpenAI Codex~\citep{openai2025codex} and Anthropic Claude Code~\citep{anthropic2026skillsdocs}---differ substantially in product surface, implementation lineage, and target workflow, yet they converge on a strikingly similar set of harness structures. That convergence is analytically significant: it suggests that the six dimensions are not incidental implementation choices but structural requirements of externalized agency. The following discussion examines these recurring patterns without tracking any single system in detail.

\paragraph{Loop and control flow.} Mature agent systems uniformly organize execution around an explicit loop that interleaves model reasoning with tool invocation and environmental observation. The harness is distinguished from the underlying model and characterized as providing the core agent loop, execution logic, and feedback pathways. Crucially, the loop includes explicit termination control---step limits, recursion depth bounds, and resource ceilings---that define the operational envelope within which the model’s reasoning unfolds.

\paragraph{Sandboxing.} Current systems implement execution isolation at different granularities. Some run each task inside a dedicated cloud sandbox with its own filesystem snapshot, network restrictions, and resource quotas; others expose graduated permission modes so that the same agent can operate at different trust levels depending on the context. These designs occupy different points in the isolation design space, but they share a common principle: sandboxing functions as a cognitive boundary that simplifies the agent’s operating environment by removing irrelevant state and restricting dangerous actions, not merely as a security perimeter.

\paragraph{Human oversight.} Rather than treating autonomy as a binary property, deployed harnesses implement configurable approval gates---hook systems that attach validation logic to specific lifecycle events such as tool invocation, file write, or subagent spawn, and application layers that route high-risk actions through approval workflows~\citep{lazaros2026human,fernandez2026agent}. The level of autonomy becomes a parameter of the harness, adjustable per task, per tool, and per organizational policy.

\paragraph{Observability.} Production systems produce structured execution traces---logs of every model invocation, tool call, memory read/write, and decision branch---that support debugging, compliance auditing, and post-incident analysis~\citep{10.1145/3759355.3759356,zhu2026verifiability}. These traces also close internal feedback loops: failed tool calls can trigger memory writes, and patterns of repeated failures can flag skills for revision. Observability is therefore the mechanism by which the harness learns from its own operation.

\paragraph{Configuration and governance.} Deployed harnesses typically stratify configuration across multiple scopes---user, project, and organization---so that the same base agent operates under different policy regimes without changes to the model or its skill artifacts. Permissions and policies function as externalized governance: constraints that would otherwise have to be embedded in prompts are instead encoded as declarative rules enforced at runtime~\citep{lee2026structuredapproachsafetycase}.

\paragraph{Context budget.} The context window remains the scarcest shared resource in any agent system. Current harnesses actively manage it through summarization of older history, staged loading that defers detailed skill guidance until a matching task is detected, and priority-based eviction of entries whose relevance has decayed. The harness orchestrates these strategies jointly because the optimal allocation depends on the current execution phase.

\vspace{1mm}
The fact that independently developed systems converge on the same set of harness dimensions is itself instructive. It indicates that the primary design challenge of externalized agency is not eliciting better completions from a model, but arranging the operational conditions under which completions become effective interventions. Harness engineering is therefore neither a synonym for memory systems nor a rebranding of tool calling. It is the broader discipline concerned with constructing the cognitive and operational environment in which externalized modules compose into coherent agency.

\subsection{Harness as Cognitive Environment}
\label{harness: cognitive environment}

The preceding sections analyzed the harness in terms of its definition, its recurring design dimensions, and its manifestation in current systems. This final section steps back to interpret the harness at a theoretical level, asking what kind of object it is rather than how it is built.

The significance of the harness extends beyond infrastructure in the ordinary software-engineering sense. A harness does not merely support an already-formed intelligence; it shapes the effective cognition of the agent by determining the environment within which reasoning unfolds. It regulates what enters the agent’s perceptual field, what is retained across turns and sessions, which operations are callable, which actions require approval, which intermediate states are exposed for revision, and which forms of failure are detectable and recoverable. The harness therefore sets the agent’s practical cognitive boundary. What the agent can know, remember, and do is not fixed by model weights alone, but by the conditions of access, persistence, and action supplied by the surrounding system.

This claim can be situated within Norman’s account of cognitive artifacts~\citep{norman1993things}. Norman characterizes cognitive artifacts as artificial devices designed to maintain, display, or operate upon information in ways that transform cognitive performance---not merely by accelerating inner computation but by changing the structure of the task itself. A harness fits this description at system scale. It does not simply augment a model with more context or more tools; it reorganizes the representational problem the model faces. By externalizing memory, formalizing procedures, introducing explicit control points, and constraining execution, the harness converts an unbounded task into a structured environment of guided action. The model’s apparent intelligence is thereby altered not only because it has more resources, but because the cognitive workload has been redistributed across artifacts, representations, and procedures outside the model. In earlier chapters, we analyzed this redistribution dimension by dimension: memory transforms recall into retrieval (Section~\ref{memory: cognitive artifact}), skills transform procedural reconstruction into guided execution (Section~\ref{skills: cognitive artifact}), and protocols transform ad hoc interaction into structured exchange (Section~\ref{protocols: cognitive artifact}). The harness is the system-level artifact that composes these individual transformations into a single cognitive environment.

Kirsh’s account of the intelligent use of space sharpens this interpretation~\citep{kirsh1995complementary}. His central observation is that cognition is shaped by how environments are arranged: spatial and representational organization can offload search, simplify choice, and reduce internal computational burden. The harness plays an analogous role for agents. It is a cognitive niche in which information, tools, permissions, and procedures are arranged so that desirable behavior becomes easier to execute and undesirable behavior becomes harder to produce. Defaults, hooks, file boundaries, skill invocation patterns, and review gates all serve as structured regularities that narrow the space of plausible action. The agent’s competence is therefore partly an ecological achievement: it arises from being embedded in an environment whose organization channels cognition productively.

The framework of distributed cognition generalizes the point. Hutchins’s formulation rejects the view that cognition resides exclusively within an individual mind, locating cognitive processes instead across people, artifacts, representations, and coordinated practices~\citep{hutchins1995cognition}. An agent system equipped with a harness is intelligible in precisely these terms. The operative intelligence is distributed across model parameters, external memory stores, executable skills, protocol definitions, tool surfaces, monitoring systems, and the runtime constraints that govern their interaction. The harness is the medium through which this distributed system is coordinated. It is thus more accurate to describe the harness as a cognitive environment than as a mere infrastructure layer. Infrastructure is one of its manifestations; environmental structuring---the design of the conditions under which cognition unfolds---is its deeper function.

%% file: text/07_cross_cutting.tex
\section{Cross-Cutting Analysis}
\label{sec:cross_cutting}

The three externalization modules are analytically distinct, but real systems derive their power from interaction among them. Sections~\ref{sec:memory}--\ref{sec:protocols} treated memory, skills, and protocols largely in isolation; Section~\ref{sec:harness} argued that the harness unifies them. This section examines the system-level couplings that arise once the modules are placed inside a harness, asks how they manifest at the model boundary, and considers where the boundary between parametric and externalized capability should be drawn.

\subsection{Module Interaction Map}

\begin{figure}[!htbp]
    \centering
    \includegraphics[width=0.5\textwidth]{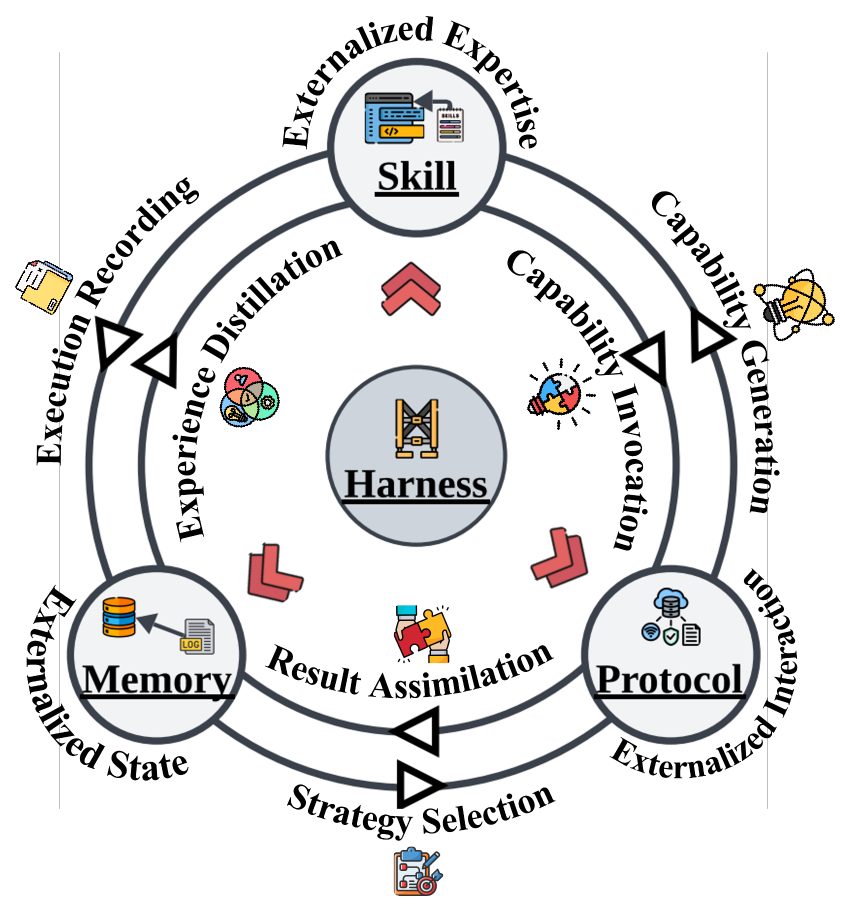}
    \caption{\textbf{Couplings among memory, skills, and protocols.} The six arrows summarize how the three externalization modules reinforce one another inside a harness. Memory supplies evidence for skill formation and protocol routing; skills turn stored experience into reusable procedures and invoke protocolized actions; protocols constrain execution and write normalized outcomes back into memory.}
    \label{fig:cut_analysis}
\end{figure}

\paragraph{Memory to skill: experience distillation.}
Repeated trajectories can be distilled into reusable procedures, making this the main path by which accumulated experience becomes codified expertise. Systems such as TED and UMEM show how episodic traces can be clustered, abstracted, and promoted into skill artifacts without modifying base-model weights~\citep{yuan2026tedtrainingfreeexperiencedistillation,ye2026umem}. Voyager makes the same logic concrete in lifelong learning: successful behaviors are retained as reusable code-level skills that can be recomposed later~\citep{wang2023voyager,zhang2025agent}.

The cross-cutting significance of this flow is that memory does not merely preserve the past; it provides the evidence from which a harness can decide what deserves to become a reusable operating pattern. The quality of the distillation step---how the system determines which trajectories generalize and which are situational---therefore conditions the reliability of the entire skill layer downstream. If distillation is too aggressive, noisy or context-dependent behaviors become entrenched as skills; if too conservative, the system fails to capitalize on hard-won experience.

\paragraph{Skill to memory: execution recording.}
The flow also runs in the opposite direction. Every skill execution generates traces, intermediate failures, and runtime refinements that would otherwise vanish with the active context window. Observability and logging infrastructure capture those trajectories as durable evidence, allowing the system to validate which skills remain reliable and which ones should be revised, split, or constrained~\citep{chen2025swe,wang2026not,wang2026museagentmultimodalreasoningagent}.

This flow is what makes the skill layer self-correcting rather than merely self-expanding. A mature skill system cannot be separated from memory management: reusable procedures only remain trustworthy if their real execution histories are continuously written back into external state. Without this recording, the harness has no empirical basis for skill maintenance, and the distillation path from memory to skill (the previous flow) operates on increasingly stale evidence.

\paragraph{Skill to protocol: capability invocation.}
Skills become operational only when they cross the boundary from abstract procedure to governed action. That transition occurs through protocols, which translate high-level intent into typed calls, lifecycle events, and permission-checked interaction surfaces~\citep{mcp_deep_dive,jsonrpc2010,hou2025model}. A skill may specify that the agent should search code, run tests, and summarize a diff, but the individual operations are carried out through protocolized interfaces to search tools, shell commands, and test runners.

The coupling matters for safety as well as for function. The OpenClaw analysis of the ``Lethal Trifecta''---sensitive data access combined with unconstrained external communication and unverified execution---illustrates that unconstrained execution remains a safety problem even when the procedural guidance itself is sound~\citep{mckerchar2026openclaw}. Protocol-level validation therefore acts as a boundary check that is independent of the skill's own correctness: even a well-written skill can be intercepted if it attempts to invoke a forbidden operation or malformed call.

\paragraph{Protocol to skill: capability generation.}
Once an interface is standardized, it becomes substantially easier to codify best practices for using it. OpenAPI and MCP do not merely make tools callable; they provide enough structural regularity for systems to package interface-specific know-how into reusable skill artifacts~\citep{openapi_spec_310,hou2025model}. The HashiCorp Agent Skills ecosystem is a concrete example: once the underlying interfaces for infrastructure management are made legible and stable through protocol contracts, domain procedures can be externalized as portable skill files rather than rederived ad hoc during each run~\citep{baghel2026hashicorp}.

This flow highlights an important asymmetry in the externalization process. Protocol standardization does not merely consume skills; it actively expands the surface on which new skills can be authored or induced. Each new stable interface is a potential seed for a family of reusable procedures. The ecosystem growth of skill artifacts therefore depends in part on the pace and quality of protocol standardization.

\paragraph{Memory to protocol: strategy selection.}
Stored context can also influence which protocol path the harness selects next. Historical success rates, user preferences, and prior failures can determine whether a request should stay local, call an external tool, or be delegated to another agent~\citep{xu2026toward,zhou2025memento}. In systems with multiple available interaction paths, memory transforms protocol selection from a static configuration into an experience-informed routing decision.

This coupling is especially visible in multi-agent settings, where the harness must choose between local execution, tool invocation via MCP, and delegation to a remote agent via A2A. If past interactions with a particular tool have consistently failed for a certain class of tasks, the routing logic can learn to prefer an alternative path. Memory therefore informs not only what the model reasons about, but which interaction channel carries that reasoning into action.

\paragraph{Protocol to memory: result assimilation.}
Finally, every protocol interaction produces state that must be preserved if it is to become part of the agent's ongoing cognition. Tool outputs, approval events, error payloads, and delegation results arrive as structured responses, often in formats richer than plain text~\citep{qin2023toolllm}. The harness must normalize these results into memory so that later reasoning can rely on verified external state rather than on reconstructed or hallucinated assumptions.

This flow closes the cycle. The protocol layer provides the evidence that memory stores, which later conditions new skill selection and new protocol routing. Without reliable result assimilation, the agent's memory becomes disconnected from its actual interaction history, and downstream flows---particularly experience distillation and strategy selection---operate on unreliable premises.

\paragraph{System-level dynamics.}
The six flows above are pairwise, but several important dynamics emerge only at the system level. First, the cycle is self-reinforcing: better memory enables better skill distillation, better skills produce richer execution traces, richer traces improve memory, and so on. This positive feedback can accelerate capability growth, but it can also amplify errors. A poisoned memory entry can lead to a flawed skill, whose execution traces further contaminate memory---a cascade that no single module's quality control can interrupt without harness-level intervention.

Second, the modules compete for the same scarce resource: the model's context window. Memory retrieval, skill loading, and protocol schemas all occupy tokens. Expanding one module's context footprint necessarily compresses the others. A harness must therefore manage not only the content of each module but also their relative budget allocation at each step of execution, a coordination problem analyzed further in Section~\ref{sec:harness}.

Third, the flows operate at different timescales. Protocol interactions are typically synchronous and fast; skill loading occurs at task or subtask boundaries; memory distillation and skill evolution unfold over sessions or longer. A harness that optimizes for one timescale---say, fast tool execution---may neglect the slower loops that determine long-term capability growth. Effective harness design requires balancing responsiveness at the fast loop with coherence at the slow loop.

\subsection{The LLM Input/Output Perspective}

Another useful viewpoint is to ask how each module manifests at the model boundary. Seen from the perspective of the context window and output surface, the harness does not simply add more components; it reorganizes what enters and leaves the model into functionally distinct layers.

\paragraph{Memory as contextual input.} Memory shapes the historical and situational input available at decision time. Instead of flooding the model with a full execution log, retrieval mechanisms select a small slice of state, prior trajectories, or entity relations that matter for the present step~\citep{du2026memoryautonomousllmagentsmechanisms}. This turns long-horizon continuity into a targeted contextualization problem and reduces context waste. The quality of this selection directly determines whether the model reasons over an accurate picture of the past or over a distorted one.

\paragraph{Skills as instructional input.} Skills shape the procedural guidance given to the model. Rather than encoding every workflow in a monolithic system prompt, the harness can load specialized instructions, examples, and constraints only when a relevant task pattern appears~\citep{jiang2026adaptationagenticaisurvey}. The model is thereby asked less often to invent a workflow from scratch and more often to interpret and follow a prepared one. The risk, discussed in Section~\ref{sec:skills}, is that overly detailed or context-consuming skill files can crowd out other inputs; the benefit is that procedural variance is reduced when the right skill is loaded at the right time.

\paragraph{Protocols as action schema.} Protocols shape the output boundary. By enforcing structured contracts such as JSON schemas, MCP messages, or OpenAPI-aligned calls, they constrain the model's generative space and make downstream execution deterministic enough to govern~\citep{hasan2026modelcontextprotocolmcp}. The output is no longer merely language to be interpreted later; it becomes a machine-readable action proposal situated inside an explicit interface. This constraint reduces the incidence of malformed tool calls and hallucinated arguments, though it also means that action expressiveness is bounded by the protocol's schema.

This input/output decomposition is analytically useful because it clarifies both the division of labor and the failure taxonomy. Retrieval errors manifest as input-selection errors: the model reasons correctly but over the wrong context. Skill failures manifest as procedural-guidance errors: the model follows instructions faithfully but the instructions themselves are flawed or mismatched. Protocol failures manifest as action-schema errors: the model's intent is sound but the output violates the interface contract. The harness makes these failure classes separable enough to debug, attribute, and optimize independently---an important property for systems where multiple modules contribute to every decision.

From a broader perspective, this tripartite organization of the model boundary---contextual input, instructional input, and action schema---can be understood as a structured form of context engineering. Rather than treating the prompt as an undifferentiated text buffer, the harness separates it into layers with distinct update rates, governance requirements, and failure modes. Each layer can be revised without disturbing the others: memory retrieval can be improved without rewriting skills, skill artifacts can be updated without changing protocol schemas, and protocol surfaces can be extended without altering memory policies. This modularity at the model boundary is one of the main practical advantages of the externalization approach.

\subsection{Parametric vs. Externalized: The Trade-off Space}

The relevant design problem is not whether intelligence should reside in the model or in the infrastructure. It is where particular burdens should live, given their update rate, reuse pattern, governance requirements, and execution cost. The following dimensions structure that partitioning decision.

\paragraph{Update frequency and temporal decay.} Fast-changing knowledge and procedures are strong candidates for externalization. APIs, organization structures, and live environment state decay too quickly to maintain reliably in model weights. Attempts to keep a model current through continual fine-tuning risk catastrophic forgetting and are often impractical at the required update frequency~\citep{cheng2024dated,qiu2025logits,zhang2025dynamic,chen2026continual}. External stores, by contrast, can be updated immediately without retraining and can maintain explicit provenance and versioning~\citep{oelen2025introducing,chinthareddy2026reliable}. Stable background capabilities---language understanding, broad reasoning, common-sense inference---decay at a much slower rate and are still more naturally carried parametrically, where they benefit from fast retrieval and deep integration with the model's representational structure.

\paragraph{Reusability and multi-agent portability.} If a capability is repeatedly needed across tasks, users, or agents, externalization improves portability and composition~\citep{tagkopoulos2025skillflow,xu2026agent,liu2025scaling}. Explicit skills, scripts, and interface artifacts can be shared, versioned, and reused across heterogeneous runtimes without requiring that each agent rediscover or retrain the same procedures. In multi-agent settings, a skill authored for one agent can be broadcast to an entire swarm, provided that the skill's assumptions about tools and protocols are met. One-off or highly idiosyncratic behavior may not justify the overhead of externalization, packaging, and maintenance~\citep{zhao2026openearth}.

\paragraph{Auditability, governance, and alignment.} Whenever inspection, approval, rollback, or policy enforcement matters, externalized artifacts have clear advantages over opaque parametric behavior~\citep{li2026security,lazaros2026human,lee2026structuredapproachsafetycase,fernandez2026agent,zhu2026verifiability}. Symbolic interfaces support circuit breakers, schema validation, and traceable execution records in a way that weights alone do not. Alignment fine-tuning (such as RLHF) provides probabilistic behavioral shaping, but externalized constraints provide deterministic enforcement at the interface level. High-stakes deployment therefore pushes the architectural boundary outward: the more consequential the agent's actions, the stronger the case for making the governing logic explicit and inspectable.

\paragraph{Latency, simplicity, and context burden.} Externalization shifts computational and organizational cost from the model's forward pass into the surrounding system. Retrieval, routing, parsing, and tool invocation all introduce latency~\citep{park2026minimizing,xu2025alignment}. Every retrieved artifact competes for limited context budget, and excessive context loading can degrade performance through information overload or the ``lost in the middle'' phenomenon~\citep{corallo2026parallel,mishra2026sok,esmi2025gpt}. For ultra-fast, low-variance, or purely semantic tasks, allowing the model to rely on its internal parametric knowledge remains substantially simpler and often more reliable.

The result is not a zero-sum contest between model intelligence and infrastructure intelligence. It is a systems-partitioning problem. Strong harnesses externalize the burdens that benefit from persistence, reuse, and control while leaving stable, fast, and generic competencies inside the model. The optimal partition is not static: as models grow more capable and as externalized infrastructure matures, the boundary will continue to shift---a dynamic explored further in Section~\ref{sec:future_frontier}.

%% file: text/08_future_discussion.tex
\section{Future Discussion}
\label{sec:future_discussion}

The preceding sections examined how memory, skills, and protocols externalize distinct cognitive burdens, and how the harness unifies them into a working agent. Those analyses describe what has already been externalized. This section asks what comes next, following the logic of externalization itself through six connected questions:

\begin{itemize}[leftmargin=1.5em]
\item Where is the boundary between parametric and externalized capability heading, and how does multi-modal perception widen that frontier?
\item Does the same logic extend from digital agents to embodied systems?
\item How can the externalization process become more autonomous?
\item What costs and risks accumulate as more is moved outward?
\item How do externalized artifacts reshape interaction at ecosystem scale?
\item How should the quality of externalization be measured?
\end{itemize}

The following subsections take up these questions in turn, moving from the shifting boundary of externalization through its embodied extension to the problem of how its benefits and costs should be assessed.

\subsection{The Expanding Frontier}
\label{sec:future_frontier}

A recurring lesson of the preceding sections is that the boundary between what stays inside the model and what gets externalized is not fixed. It shifts as models, tasks, and infrastructure co-evolve. Understanding that boundary---and anticipating where it will move next---is therefore a central design question for agent systems.

In one direction, model improvement can pull capability back inward. A model that reliably produces structured output needs less format validation in the harness; one with a larger effective context window may tolerate simpler memory architectures; one with stronger intrinsic tool-use ability may require less elaborate intent-capture logic. Each such advance renders some piece of external infrastructure redundant. In the opposite direction, richer harnesses create new demands on models: operating inside a structured runtime requires respecting schemas, cooperating with permission checks, and coordinating with staged context injection~\citep{zhang2025dynamic,cheng2024dated}. The frontier therefore moves in both directions at once, and a central engineering challenge is knowing when to externalize further and when to retract.

Within this shifting landscape, several classes of cognitive work that today remain largely implicit are plausible candidates for further externalization.

\paragraph{Planning and goal management.}
Current agents typically generate plans through in-context reasoning, producing decompositions that are ephemeral---they exist only in the active generation and are lost once the context resets. Early agent frameworks such as BabyAGI already experimented with persistent task queues~\citep{nakajima2023babyagi}, and file-centric state abstractions like InfiAgent materialize planning artifacts outside the prompt~\citep{yu2026infiagentinfinitehorizonframeworkgeneralpurpose}. The direction points toward plans as first-class harness objects: persistent, inspectable, revisable, and shareable across agents or between agents and humans. That would convert planning from a transient reasoning act into a managed state artifact---the same representational shift that memory already performs for historical context.

\paragraph{Evaluation and verification.}
Most evaluation logic today lives either inside the model's chain of thought or in external benchmark harnesses that run post hoc. Externalizing evaluation criteria, rubrics, and verification procedures as runtime harness components---rather than leaving them implicit in model judgment---would allow the agent to check its own outputs against explicit standards during execution. Early signs of this direction are visible in verifiability-first engineering frameworks~\citep{zhu2026verifiability} and in self-refine loops that separate generation from critique~\citep{madaan2023selfrefineiterativerefinementselffeedback}. The broader opportunity is to treat evaluation as externalized quality infrastructure rather than as a post-hoc measurement.

\paragraph{Orchestration logic itself.}
The most recursive form of externalization is making the harness's own configuration, policies, and execution logic into objects that the agent can inspect, critique, and revise. Once orchestration logic is externalized, the agent system can adapt not only what it knows and does, but how it organizes knowing and doing. This direction connects directly to the next subsection.

\paragraph{Multi-modal externalization.}
The externalization framework developed so far assumes text as the dominant representational medium: memory stores textual traces, skills encode natural-language procedures, and protocols exchange structured text messages. As foundation models become natively multi-modal---processing images, video, audio, and screen content alongside text---each externalization dimension faces new design demands. Multi-modal skills must encode not only textual procedures but also visual perception workflows and cross-modal decision logic; early examples include computer-use skills that package GUI interaction sequences as reusable units~\citep{chen2026cua}. Multi-modal memory must index and retrieve visual and auditory experience, not only text-based episodic traces; MemVerse, for instance, maintains a multimodal knowledge graph that periodically distills fragmented sensory experience into more abstract representations~\citep{liu2025memverse}, and MuSEAgent accumulates stateful multimodal experiences to inform future reasoning~\citep{wang2026museagentmultimodalreasoningagent}. Multi-modal reasoning distillation extends the skill-acquisition loop to non-textual modalities: TED demonstrates that successful multimodal reasoning trajectories can be distilled into reusable experience without additional training~\citep{yuan2026tedtrainingfreeexperiencedistillation}. The broader implication is that multi-modal externalization is not simply a matter of adding new data types to existing stores. It changes the design assumptions of skill specification, memory indexing, and protocol schemas, and it opens a substantially wider frontier for the externalization of cognitive burden~\cite{wang2026oscar,xu2026photobench}.

\subsection{From Digital Agents to Embodied Externalization}
\label{sec:future_embodied}

The externalization framework developed in this paper applies to digital agents that read, write, and call APIs. A natural question is whether the same architectural logic extends to embodied systems---robots that must also perceive, move, and physically interact with the world. Recent developments in robot learning suggest that it does, and that the embodied domain is undergoing a decomposition strikingly parallel to the one analyzed here.

\paragraph{The monolithic starting point.}
Early approaches to embodied intelligence pursued an end-to-end strategy analogous to the pre-externalization LLM agent. Vision-Language-Action (VLA) models~\citep{brohan2023rt2,kim2024openvla} were positioned as monolithic ``brains'': given a natural-language instruction and a visual observation, the model directly outputs a continuous action sequence, handling perception, reasoning, planning, and motor control within a single forward pass. This design mirrors the pattern in which early LLM agents attempted to manage memory, skills, and orchestration entirely through in-context reasoning---and it encountered the same category of limitations. Complex multi-step tasks exceeded the model's planning horizon; failures in intermediate steps could not be diagnosed or recovered from; and the tight coupling of high-level cognition with low-latency motor control created irreconcilable requirements on inference speed and model capacity.

\paragraph{Decomposition: the cerebrum--cerebellum split.}
The emerging architectural response recapitulates the externalization logic at the level of the whole body. A high-level \emph{robot agent}---typically an LLM or multimodal model---assumes the role of cerebrum: it interprets goals, decomposes tasks into subtask sequences, maintains state across steps, handles exceptions, and revises plans when execution feedback indicates failure~\citep{ahn2022saycan,singh2023progprompt,liang2023code}. VLA models, meanwhile, are repositioned as a \emph{cerebellum}: each one becomes a callable skill module responsible for a single atomic manipulation primitive---grasping, placing, pouring, inserting---executed with real-time sensorimotor feedback and low-latency control. The VLA no longer decides \emph{what} to do; it ensures that \emph{how} it is done is precise, stable, and adaptive to local physical perturbations.

This decomposition maps directly onto the externalization dimensions of the present paper. Task planning and goal management migrate from the VLA's implicit parametric reasoning into an explicit, inspectable agent loop---precisely the shift from in-context planning to externalized plan objects discussed in Section~\ref{sec:future_frontier}. Each VLA skill module functions as an externalized skill artifact: a reusable, composable unit with a defined interface, analogous to the skill files and tool specifications analyzed in Section~\ref{sec:skills}. The communication between agent and skill---structured action requests, execution status reports, error codes---constitutes a protocol layer that enables the agent to orchestrate heterogeneous motor capabilities without embedding their implementation details.

\paragraph{Why the parallel matters.}
The convergence is not coincidental. Both digital and embodied agents face the same fundamental tension: a single model cannot simultaneously optimize for slow, deliberative cognition and fast, reactive execution. Externalization resolves this tension by routing each class of cognitive work to the substrate best suited for it---persistent, inspectable structures for planning and memory; specialized, low-latency modules for execution. In the digital case the execution modules are tool calls and code interpreters; in the embodied case they are visuomotor policies. The harness pattern---a runtime that loads context, dispatches skills, enforces protocols, and manages state---is equally applicable to both, suggesting that embodied and digital agent architectures may ultimately share not only a design philosophy but a concrete engineering stack.

\paragraph{Open challenges.}
Embodied externalization introduces constraints that the digital case does not face. Physical actions are irreversible in ways that API calls are not: a dropped object cannot be ``rolled back.'' Real-time control demands latency budgets orders of magnitude tighter than text generation. Perception is noisy, and the gap between simulated training environments and physical deployment remains substantial. These constraints will shape how memory, skills, and protocols are designed for embodied harnesses, but they do not change the fundamental argument: the logic of externalization---decomposing monolithic capability into specialized, composable, and governable external structures---extends naturally from digital cognition to physical action.

\subsection{Toward Self-Evolving Harnesses}
\label{sec:future_self_evolving}

Most current agent systems still rely on humans to revise memory policies, rewrite skill artifacts, and tighten execution logic after failures. If orchestration logic is itself externalized---as the previous subsection suggests---then the harness becomes an object that can be adapted programmatically rather than only manually. The question is how to make that adaptation reliable.

From a systems perspective, self-evolution can occur at three levels. At the \emph{module level}, the architecture stays fixed but internal policies---retrieval granularity, skill-ranking heuristics, protocol-routing rules---are adjusted in response to observed failures. At the \emph{system level}, the execution pipeline itself is restructured: scheduling strategies, execution order, or resource allocation may change when logs reveal recurring bottlenecks that local tuning cannot resolve. At the \emph{boundary level}, the scope of the harness expands or contracts as models and tasks change, adding new externalized components where needed and pruning redundant ones---precisely the frontier dynamics discussed in Section~\ref{sec:future_frontier}.

Several technical pathways are emerging. Reinforcement learning can optimize discrete runtime policies---search depth, compression ratio, retry strategy---against rewards such as task success, latency, or resource cost. Program synthesis treats harness adaptation as code repair: the model proposes patches after a failed trajectory, and sandboxed tests validate them before deployment. Evolutionary methods search over the topology of the harness---how modules are connected and in what order they are invoked. Imitation learning provides a stronger prior when exploration is too costly, by distilling execution logs from human experts or strong models into better orchestration patterns. These pathways target different search spaces---policy, program, structure, and prior experience---and are likely to be combined rather than used in isolation.

Self-evolution is attractive because it targets infrastructural failure modes directly, but it also amplifies the costs and risks discussed next: an adaptive harness that drifts without adequate governance can introduce new failure modes faster than it resolves old ones.

\subsection{Costs, Risks, and Governance}
\label{sec:future_costs}

As more cognitive burden is moved outward, two classes of cost accumulate: cognitive overhead from the externalized infrastructure itself, and security risks from the expanded attack surface.

\paragraph{Cognitive overhead.}
Externalization is not free~\citep{wang2026beyond}. Every additional memory layer, API schema, or safety rule imposes latency and reasoning overhead, and past a certain point the model spends more effort discovering, parsing, and coordinating modules than solving the task itself. In memory, over-retrieval floods the context with marginally relevant traces. In skills, verbose or overlapping files compete for context budget and can cause the model to follow local procedure while losing sight of the global objective. In protocols, tool sprawl turns action selection into an unnecessary disambiguation problem.

These failure modes suggest that the design target should be efficient and utility-positive rather than maximal externalization~\cite{liu2025real}. \emph{Minimal sufficiency} asks whether a given module actually reduces the model's cognitive burden or merely adds one. \emph{Lazy loading} defers detailed guidance until the task structure requires it. \emph{Budget-aware routing} treats context allocation as an explicit optimization variable, dynamically adjusting how much space is devoted to memory, skills, and protocol metadata as the task phase changes~\citep{zhang2026learning,patel2025dynamic,sui2026act}. A good harness simplifies the model's decision problem; it does not create a second one.

\paragraph{Security and integrity risks.}
Cognitive overhead is a performance cost; the security dimension is more consequential. Once cognitive and procedural burdens are relocated into external artifacts, those artifacts become targets---and the threats map directly onto the three harness dimensions. Memory poisoning can silently distort future reasoning through corrupted episodic traces or factual stores. Malicious skill injection can embed adversarial procedures into the agent's reusable repertoire. Protocol spoofing---forged tool manifests or manipulated endpoints---can cause unauthorized actions under the appearance of legitimate interaction~\citep{liu2026agent,guo2026skillprobe,wang2026skills,lin2025superplatforms}. These risks are compounded when externalization becomes self-evolving (Section~\ref{sec:future_self_evolving}): adapting to new tasks can degrade old ones, accumulated patches can obscure system behavior, and optimization targets can be distorted when human supervision weakens.

\paragraph{Governance as infrastructure.}
The implication is that externalization must be paired with governance---not as an afterthought, but as a co-designed layer of the harness. Mandatory review gates for critical updates, provenance tracking for memory and skill changes, deterministic rollback mechanisms, and regression testing all become part of the infrastructure. The quality of an externalized system is therefore measured not only by what it enables, but by how transparently and reversibly it does so. This criterion also informs evaluation, as discussed in Section~\ref{sec:future_eval}.

\subsection{From Private Scaffolding to Shared Infrastructure}
\label{sec:future_shared}

The externalization described so far is largely agent-centric: memory serves one agent's continuity, skills are loaded as local packages, and protocols often remain framework-bound. As collaboration chains lengthen, however, externalization begins to shift from private scaffolding toward shared infrastructure~\citep{wang2025mirix,li2026organizing,nie2026holoswebscalellmbasedmultiagent}. This changes the unit of analysis from the individual agent to the ecosystem.

\paragraph{Shared artifacts.}
The clearest sign is the emergence of shareable artifacts across all three dimensions. Shared memory shifts the question from ``what I remember'' to ``what we know,'' turning memory into a transactive system of shared state, indices, and common ground~\citep{wegner1987transactive,zhao2026shardmemo}. Shared skills turn procedural expertise into public capability units that can be reused, forked, and maintained across agents~\citep{ling2026agent}. Shared protocols provide the common grammar that makes such coordination interoperable across platforms and organizations~\citep{yang2025survey}.

\paragraph{Division of labor and collective learning.}
Once these structures are shared, agent systems can differentiate roles rather than replicate the same full stack everywhere. Drawing on stigmergy~\citep{theraulaz1999brief}, failure trajectories can accumulate in shared memory while successful paths crystallize into shared skills. Learning then diffuses through external structures rather than only through joint parametric training.

\paragraph{Institutionalization and its tensions.}
As memory schemas, skill specifications, and protocol bindings are repeatedly validated, they begin to function less like temporary scaffolding and more like institutions: shared operating procedures and standards that coordinate behavior at ecosystem scale~\citep{hutchins1995cognition}. But shared infrastructure also introduces new governance problems~\citep{deng2025ai,liu2026agent,kong2025survey}. Infrastructure drift, malicious or low-quality artifacts, and premature or delayed standardization can all destabilize the ecosystem~\citep{guo2026skillprobe,timmermans2010world}. The governance costs identified in Section~\ref{sec:future_costs} are therefore amplified when externalization becomes collective: version control, permission auditing, provenance, and rollback become part of the institutional design of agent systems, not just the engineering of individual harnesses.

\subsection{Measuring Externalization}
\label{sec:future_eval}

Most current benchmarks evaluate agents primarily through task completion under fixed prompts and fixed model settings~\citep{zhu2025evolutionary,mishra2026sok}. That is useful for comparing base-model capability, but it systematically under-measures the contribution of externalized infrastructure. A harness that improves reliability through better memory retrieval, more precise skill loading, or tighter execution governance will show up only as a higher pass rate, with no way to attribute the gain to its actual source.

A richer evaluation agenda would assess the quality of externalization along dimensions that current benchmarks largely ignore. \emph{Transferability} asks whether the same harness configuration maintains its effectiveness when the underlying model is swapped---a direct test of how much capability resides in external infrastructure versus weights. \emph{Maintainability} measures how gracefully the system degrades when skills, memory policies, or protocol schemas are updated. \emph{Recovery robustness} tests whether the agent can detect failures, roll back partial actions, and resume from checkpoints. \emph{Context efficiency} quantifies how much of the context budget is consumed by harness overhead versus task-relevant reasoning. \emph{Governance quality} evaluates whether the externalized system meets the transparency and reversibility requirements identified in Section~\ref{sec:future_costs}.

Concrete evaluation strategies might include ablation studies that remove individual harness components and measure the resulting degradation; cross-model transfer tests that hold the harness constant while varying the base model; and long-horizon reliability metrics that track success rates, cost, and drift over extended multi-session interactions rather than single-turn completions. Until such methods mature, the field will continue to attribute to model intelligence what is partly an achievement of externalization design.
For instance, the Agent Humanization Benchmark (AHB) suggests that agent evaluation should extend beyond task completion to the humanization of observable behavior at the user-interface boundary, especially for mobile GUI agents operating in human-centric environments~\cite{zhu2026turing}.

\vspace{2mm}
\noindent Taken together, these six directions trace the continuing logic of externalization beyond its current state. The frontier is expanding as new cognitive burdens---including multi-modal perception and cross-modal reasoning---become candidates for externalization; the same decomposition logic is extending from digital agents to embodied systems, where the cerebrum--cerebellum split recapitulates the separation of planning from execution; the process is becoming more autonomous through self-evolving harnesses; the trade-offs are sharpening as cognitive overhead and security risks accumulate; the scope is widening from private scaffolding to shared infrastructure; and the evaluation challenge is growing more pressing as externalization's contribution remains invisible to model-centric benchmarks. The common thread is that externalization is not a one-time architectural decision but an ongoing design process whose boundaries, mechanisms, costs, and quality criteria co-evolve with the models and ecosystems they serve.

%% file: text/09_conclusion.tex
\section{Conclusion}
\label{sec:conclusion}

This paper has argued that externalization is the transition logic connecting many of the most important developments in LLM agents. Reliable agency increasingly depends on relocating selected cognitive burdens out of the model and into explicit infrastructure: memory externalizes state across time, skills externalize procedural expertise, protocols externalize interaction structure, and the harness coordinates these layers into a working runtime.

From this perspective, the move from weights to context to harness is not just a sequence of engineering tricks. It marks a shift in where agent capability is organized. Some burdens remain well handled parametrically, but others become more reliable once they are made persistent, inspectable, reusable, and governable outside the model.

What unifies these forms of externalization is representational transformation. Memory turns recall into retrieval, skills turn improvised generation into guided composition, and protocols turn ad hoc coordination into structured exchange. The effect is not simply to add more components around the model, but to change the task the model is being asked to solve.

This reframing also clarifies the agenda ahead. The key questions are no longer only how to build stronger models, but how to partition capability between models and infrastructure, how to evaluate the contribution of externalized systems, and how to govern the shared artifacts on which agents increasingly rely.

The broader implication is that progress in agents will come from the co-evolution of models and external infrastructure rather than from either in isolation. On that view, better agents are not merely better reasoners. They are better organized cognitive systems.